\documentclass[aps,floats,floatfix,nofootinbib,superscriptaddress,12pt]{revtex4}
\pdfoutput=1
\usepackage{graphicx}
\usepackage{amsmath}    
 
\usepackage{color}
\def\be{\begin{equation}}
\def\ee{\end{equation}}
\def\bea{\begin{eqnarray}}
\def\eea{\end{eqnarray}}

\newcommand{\lsim}{ \mathop{}_{\textstyle \sim}^{\textstyle <} }

\newcommand{\ev}{{\rm eV}}
\newcommand{\kev}{{\rm keV}}

\newcommand{\gev}{{\rm GeV}}

\newcommand{\snu}{{\tilde \nu}}
\newcommand{\neutone}{{{\tilde \chi}_1^0}}
\newcommand{\neuttwo}{{{\tilde \chi}_2^0}}
\newcommand{\slepr}{{{\tilde l}_R}}

\begin{document}

\title{Mixed Sneutrinos, Dark Matter and the LHC}

\author{Zachary Thomas}
\affiliation{Department of Physics,Williams College,\\ Williamstown, MA 01267}
\author{David Tucker-Smith}
\affiliation{Department of Physics,Williams College,\\ Williamstown, MA 01267}
\author{Neal Weiner}
\affiliation{Center for Cosmology and Particle Physics, \\
  Dept. of Physics, New York University, \\
New York, NY 10003}

\begin{abstract}

We study the  phenomenology of supersymmetric models in which gauge-singlet scalars mix 
with the MSSM sneutrinos through weak-scale $A$ terms.  After reviewing the constraints on 
mixed-sneutrino dark matter from measurements of $\Omega_{CDM}$ and from direct-detection
experiments, we explore mixed-sneutrino signatures relevant to the LHC.  
For a mixed-sneutrino
LSP and a right-handed slepton NLSP, decays of the lightest neturalino can produce opposite-sign, 
same-flavor (OSSF) dileptons with an invariant-mass distribution shifted away from the kinematic
endpoint.  In different parameter regions, the charginos and neutralinos produced in cascades all
decay dominantly to the lighter sneutrinos, leading to a kinematic edge in the jet-lepton invariant-mass
distribution from the decay chain $\tilde{q} \rightarrow \chi^- q \rightarrow \snu^* l q$, without an OSSF dilepton signature.  We explore the possibility of using mass estimation methods to distinguish this mixed-sneutrino jet-lepton signature
from an MSSM one. 
Finally, we consider signatures  associated with Higgs-lepton or $Z$-lepton production
 in cascades involving the heavier sneutrinos.

\end{abstract}

\maketitle

\section{Introduction}

The overwhelming evidence that the majority of the matter in the universe is non-baryonic compels us to consider extensions of the standard model which include new fields that are electrically neutral. As was pointed out by Goodman and Witten \cite{Goodman:1984dc} fields at the weak scale naturally yield the appropriate relic abundance to explain the observed density of dark matter in the universe. Taken alone, this may be the strongest motivation for new physics at the weak scale.

A second element of physics beyond the standard model that has become firmly established in recent years is  neutrino mass. Although it is usually assumed that neutrino mass is generated at very short distances, thus explaining its smallness through the seesaw mechanism \cite{Yanagida, GellMann:1980vs}, this hypothesis remains untested, 
motivating us to consider alternative possibilities.

For example, in supersymmetric theories with right-handed neutrinos at or below the weak scale, small Dirac neutrino masses can be generated as a supersymmetry-breaking effect, or else small Majorana neutrino masses can be generated radiatively or through a weak-scale seesaw \cite{ArkaniHamed:2000bq,ArkaniHamed:2000kj,Borzumati:2000mc}.  Moreover, if the scalar partners of the right-handed neutrinos mix appreciably with the MSSM sneutrinos, the lightest ``mixed'' sneutrino  can be considerably lighter than the Z boson \cite{ArkaniHamed:2000bq,Borzumati:2000mc,Chou:2000cy}, and can also be a viable dark matter candidate \cite{ArkaniHamed:2000bq}.

The existence of new states beyond those of the MSSM may be crucial for searches at the LHC. Unlike in previous experiments, new particles are likely to be produced in bunches in potentially long cascades. The presence of new states in the cascade chains can lead to new signatures and remove expected ones. It is an intriguing and important question to what extent analyses that can be performed at the LHC might give some indication that a mixed sneutrino is present in the spectrum of the theory.

In this paper we study the cosmology and potential LHC signatures associated with mixed sneutrinos.  
In section \ref{sec:dm}, we consider the possibility of mixed-sneutrino dark matter, calculating the relic abundance for a range in parameters, and imposing  all present constraints from direct detection experiments, including the most recent from XENON \cite{Angle:2007uj}.   In agreement with \cite{Arina:2007tm}, we find that mixed-sneutrino dark matter is viable over substantial parameters regions.  We consider the extent to which lepton number violation in the sneutrino mass matrix might suppress rates at direct detection experiments, and discuss the connection to neutrino masses.

In section \ref{sec:lhc} we explore  the collider phenomenology of mixed-sneutrinos.  
The first signature we consider, which persists even for very small sneutrino mixing angles, arises from leptonic decays of the lightest neutralino in the case where the LSP is the lightest sneutrino and the NLSP is a right-handed slepton.  The possibility of lepton production from the decays of the lightest neutralino was also pointed out in ref.~\cite{Covi:2007xj}, which studied the phenomenology of a  sneutrino NLSP with a gravitino LSP.   
We study the invariant-mass distribution of opposite-sign, same flavor dileptons from these decays and find that it is shifted away from the kinematic endpoint.  The other signatures we consider require somewhat larger mixing angles, as they involve decays of non-NLSP superpartners straight to the lightest sneutrino.  For example, in a broad region of parameter space, the gauginos produced in cascades decay almost exclusively directly to the lightest sneutrinos.  In this case one has a kinematic edge in the lepton-jet invariant mass distribution, 
without a corresponding dilepton edge.  We discuss the possibility of distinguishing this signature from MSSM ones, for example
by using recently proposed methods to estimate the masses involved in cascade decays.  Finally, we consider signatures   from Higgs-lepton or Z-lepton production in cascades involving the heavier sneutrinos, which can lead to distinctive $bbl$, $\gamma \gamma l$ or trilepton invariant mass distributions.

\section{Mixed Sneutrino Dark Matter}
\label{sec:dm}
To be viable, WIMP dark matter candidates must pass three essential tests. First, they must be neutral, both to allow early growth of structure and to have evaded detection. Second, their relic abundance must match the measured value of the dark matter energy density,  $\Omega_{CDM} h^2 \sim 0.1$ \cite{Spergel:2006hy}. Third, given the appropriate relic density they must evade direct-detection experimental limits, the most severe of which  presently come from XENON \cite{Angle:2007uj} and CDMS \cite{Akerib:2005kh}.

The sneutrino was long ago considered an intriguing dark matter candidate \cite{Hagelin:1984wv,Ibanez:1983kw}, but is no longer viable as it fails the combined relic abundance and direct detection requirements. In particular, a light sneutrino with the appropriate relic abundance would significantly modify the invisible $Z$-width, in conflict with observation. A heavy sneutrino must be of the order 600 GeV to achieve the correct relic abundance \cite{Falk:1994es}, and even then is in clear conflict with direct detection experiments. Similarly, even if a moderate-mass ($ \sim100$ GeV) sneutrino had come out with the correct relic abundance, it would have been seen at experiments such as CDMS and XENON. One proposal  for saving  sneutrino dark matter is to  suppress coannihilation of the sneutrino's scalar and pseudo-scalar components by making them non-degenerate \cite{Hall:1997ah}, which also eliminates direct detection constraints arising from  $Z$-exchange contributions to the scattering of sneutrinos off of nuclei. Unfortunately, this scenario implies a $\nu_\tau$ mass well above the experimental limit.

The problems with sneutrino dark matter mainly stem from the large coupling of sneutrinos to Z bosons.   However, because the sneutrino is neutral under electromagnetism, it is free to mix with any additional neutral scalar field, assuming that the field carries lepton number or that lepton number is not a good symmetry of the low-energy theory.  This possibility was explored in \cite{ArkaniHamed:2000bq,Borzumati:2000mc,Chou:2000cy}.  The mixing suppresses the coupling of the lightest sneutrino to the Z, and its mass  is allowed to be less than $m_Z/2$ for mixing angles satisfying $\sin \theta \lsim 0.4$. Because the sneutrino annihilation rate in the early universe is also suppressed,  the appropriate relic abundance can be achieved \cite{ArkaniHamed:2000bq}.

Related scenarios for sneutrino dark matter include non-thermal right-handed-sneutrino dark matter (where the mixing is extremely tiny) \cite{Asaka:2005cn,Gopalakrishna:2006kr}, and thermally produced right-handed-sneutrino dark matter
in the presence of an extra $U(1)$ \cite{Lee:2007mt}.

The outline for the rest of the section is as follows.  First,
 we review models of mixed-sneutrino dark matter.  Then we discuss the relic abundance calculation and identify  cosmologically preferred parameter regions. With these results in mind, we  review constraints from direct-detection experiments, and find, in  agreement with \cite{Arina:2007tm}, that significant regions of parameter space remain viable. 
Lepton-number violation in the sneutrino mass  matrix can suppress the scattering of sneutrinos off of nuclei, and
thus direct-detection rates,  but it also radiatively generates neutrino masses that tend to be beyond experimental limits. We discuss a few scenarios in which these neutrino masses are not problematic, and then briefly consider the implications of sneutrino dark-matter  for neutrino telescope indirect detection experiments. Finally, we comment on scenarios in which the gravitino is the LSP, with a mixed-sneutrino NLSP.

\subsection{Mixed Sneutrinos with Large or Small Yukawas}
\label{sec:yuk}
The model we consider is quite simple.
To the  MSSM, we add one or more additional standard-model-singlet superfields $N_i$, with supersymmetry-breaking trilinear couplings of the form $A_{ij} \tilde n_i \tilde l_j h_u$. Restricting ourselves for the moment to one generation, this leads to a mass matrix of the form
\be
M^2_{\tilde \nu} =
\begin{pmatrix}
m_L^2 + \frac{1}{2}m_Z^2 \cos 2\beta & A v \sin \beta \\ A v \sin \beta & m_{\tilde n}^2
\end{pmatrix},
\label{eq:matrix}
\ee
with mass eigenstates $\tilde \nu_1 = \cos \theta \tilde n^* -\sin \theta \tilde \nu$ and $\tilde \nu_2 = \sin \theta \tilde n^* +\cos \theta \tilde \nu$. Motivated by the possibility of mixed-sneutrino dark matter, we take $\sin^2 \theta < 0.5$, so the lighter state is more singlet than active sneutrino. If this lighter state is heavier than $m_Z/2$ there are no immediate constraints on $\sin \theta$, while if it is lighter than $m_Z/2$, the $Z$-width constraint requires $\sin \theta < 0.4$.

The $A$ terms for the superpartners of the standard model fermions are typically thought to be related to the associated  Yukawa couplings.  Given the apparent smallness of the neutrino Yukawa couplings, one thus might not expect sizeable mixing between the active and sterile sneutrinos.  However, it is possible that bare Yukawa couplings for the neutrinos are forbidden by a $U(1)_n\otimes U(1)_l $ symmetry that acts independently on the singlet and lepton-doublet superfields.  If this symmetry is broken only by supersymmetry-breaking fields, then weak-scale $A$ terms and tiny Yukawa couplings are perfectly compatible.

Moreover, one can instead have large Yukawa couplings and still have massless neutrinos, as we now describe.
If  the fields $N_i$ come with fields $\bar N_i$ which carry opposite lepton number charge, we can consider the following superpotential,
\be
W \supset \lambda N L H_u + m_N N \bar N.
\ee
When the Higgs acquires an expectation value, there are Dirac masses between $\nu$ and $n$, as well as between $n$ and $\bar n$. Because of the mismatch between states with lepton number $+1$ and $-1$, there is a massless state in the theory. This is essentially the same mechanism that keeps the neutrino light in the standard model. If $m_N > \lambda v_u$ the massless state will then be mostly standard model-neutrino. Constraints on this scenario come from a variety of precision electroweak measurements, principally from measurements of the couplings of  charged leptons to neutrinos. For light ($m_N \lsim m_Z$) neutrinos, $\lambda v$ should be smaller than about $m_\tau$. However, for heavier neutrinos, a larger Yukawa is allowed, even for much lighter sneutrinos. This setup results in a $3\times 3$ sneutrino mass matrix instead of the $2\times 2$ one  of eqn.~(\ref{eq:matrix}),
\be
\begin{pmatrix}
m_L^2 + \frac{1}{2}m_Z^2 \cos 2\beta &A v \sin \beta + \lambda \mu v \cot \beta & \lambda m_N v \tan \beta \\ A v \sin \beta + \lambda \mu v \cot \beta & m_{\tilde n}^2 & 0 \\ \lambda m_N v \tan \beta & 0 &  \bar m_{\tilde n}^2
\end{pmatrix}.
\ee
In the limit where $m_N$ (and thus $ \bar m_{\tilde n}^2$) are very large, the effective mass matrix for the lighter sneutrinos is the same as in eqn.~(\ref{eq:matrix}) except  with the replacement $A \rightarrow X$, where $X = A + \lambda \mu \cot \beta$.

In this setup, non-zero neutrino mass can be generated through higher-dimension operators, or radiatively if a small lepton-number-violating terms appear in the
full sneutrino mass matrix. 
For our calculations of direct detection rates and relic abundances we will restrict ourselves to the model with negligible Yukawa couplings, and leave a thorough analysis of the relic abundance of the Yukawa model to future work. However, it is worth noting that even within the mixed-sneutrino framework, there is great room for variation.

\subsection{Relic Abundance of Mixed Sneutrinos}
\begin{figure}[htbp] 
   \centering
   \includegraphics[width=2in]{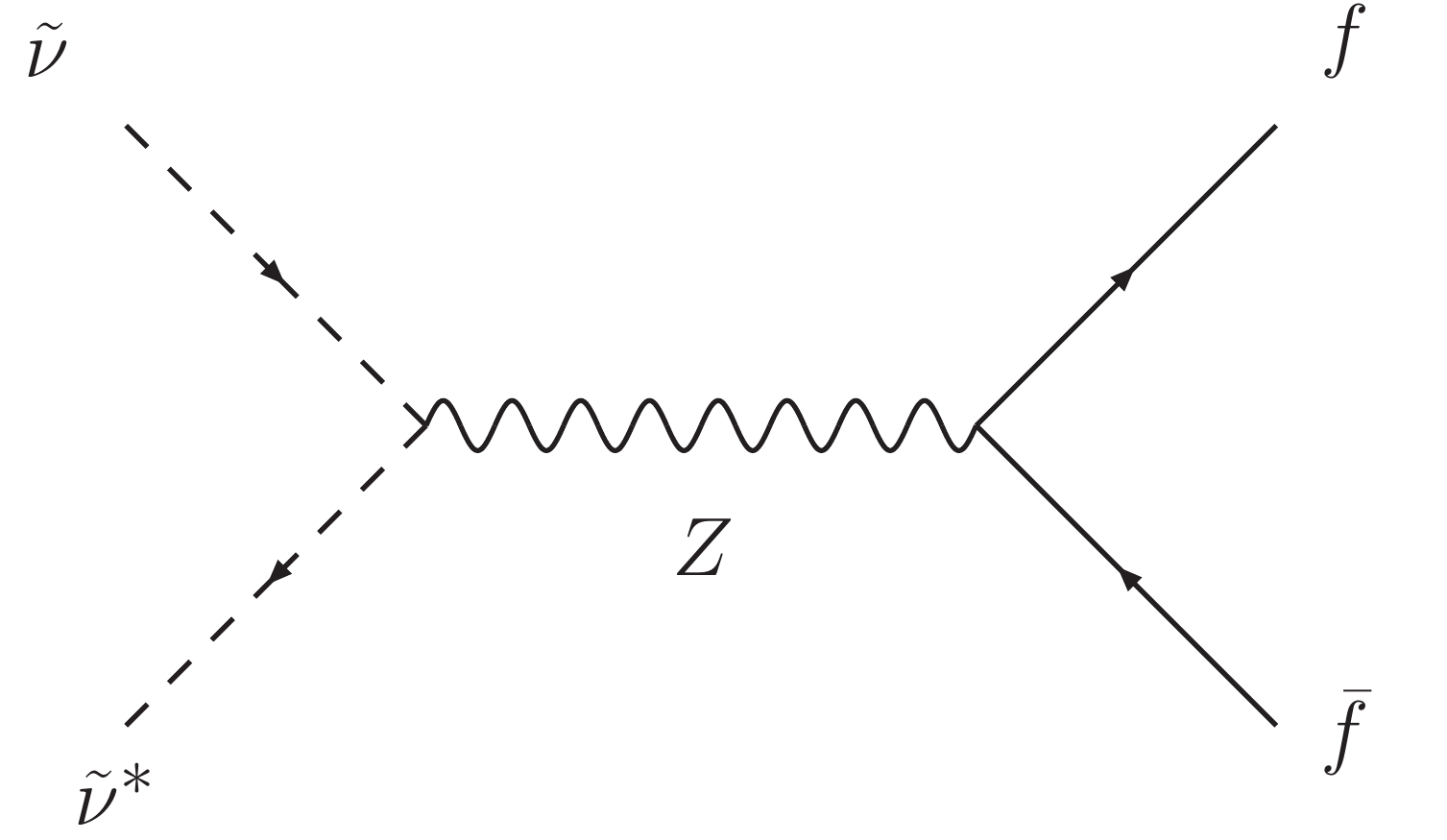} 
      \includegraphics[width=2in]{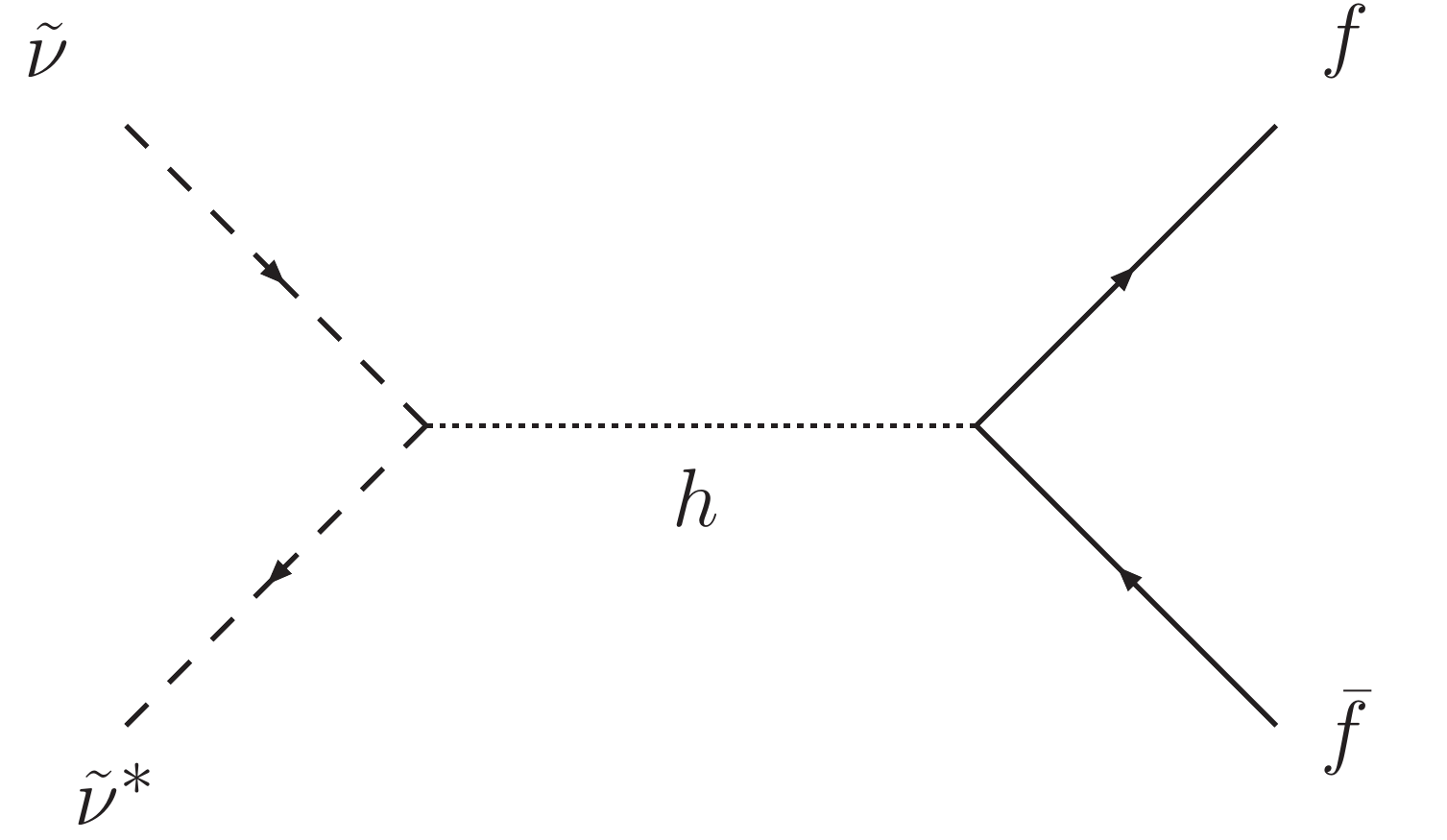} 
         \includegraphics[width=1.2in]{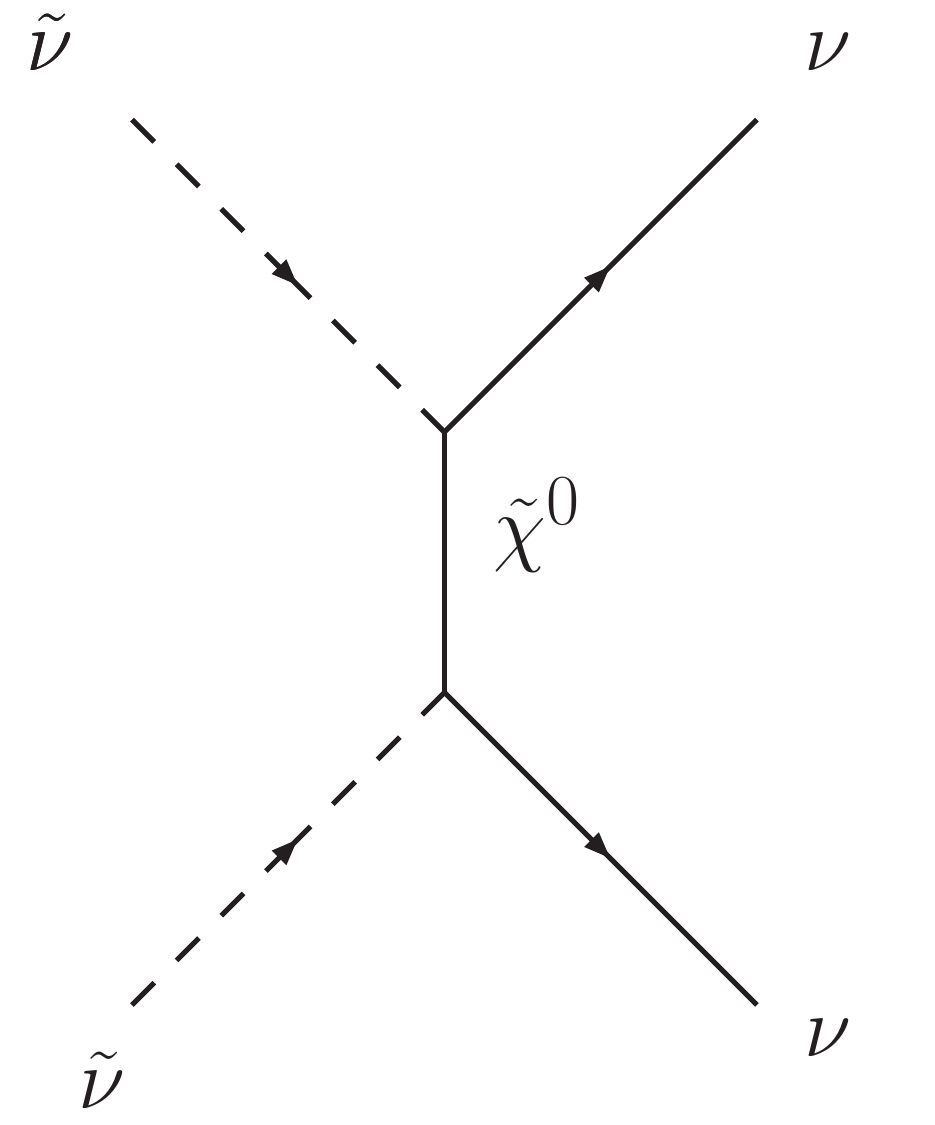} 
\caption{Diagrams contributing to the mixed-sneutrino annihilation rate in the early universe.}
  \label{fig:anndiag}
\end{figure}

The dominant annihilation channels for mixed sneutrinos in the early universe are shown in figure~\ref{fig:anndiag}.  These
include $s$-channel $Z$ exchange, $t$-channel neutralino exchange (to $\nu \nu$ or $\nu \bar \nu$), and $s$-channel Higgs exchange (to fermions, or, for heavier sneutrinos, to gauge bosons and Higgs bosons). The contribution from Higgs exchange is enhanced by the large $A$-terms, and is often dominant.  

To calculate the relic abundance of the mixed-sneutrino LSP, we use the micrOMEGAs 2.0 code \cite{Belanger:2006is}
with the MSSM model files modified to incorporate the mixed sneutrino. Superpartner and Higgs particle spectra are calculated using SuSpect  \cite{Djouadi:2002ze}. We assume that only a single mixed sneutrino
has a significant relic abundance today.   Even if multiple right-handed sneutrinos are appreciably mixed with the active ones, this will still be
true provided the light sneutrinos are not highly degenerate.

In our calculations we fix the values of the MSSM parameters at the weak scale.  We take the input parameters in the sneutrino sector to be the mixing angle $\theta$, the LSP mass $m_{\tilde \nu_1}$, and the soft mass-squared for the left-handed sleptons, $m_L^2$.  Once $\theta$ and $m_{\tilde \nu_1}$ are fixed, both the relic abundance and direct-detection rate (discussed in the following section) are both quite sensitive to $m_L^2$.  This is because increasing $m_L^2$ increases   the $A$ (or $X$)-parameter, thereby enhancing the annihilation rate via $s$-channel Higgs exchange and the cross-section for Higgs-mediated $\tilde \nu_1$--nucleon scattering.  The gaugino masses $M_1$ and
$M_2$ can also be important in determining the relic abundance, as  $t$-channel  neutralino exchange is another potentially significant
annihilation channel for $\tilde \nu_1$.

\begin{figure}[]
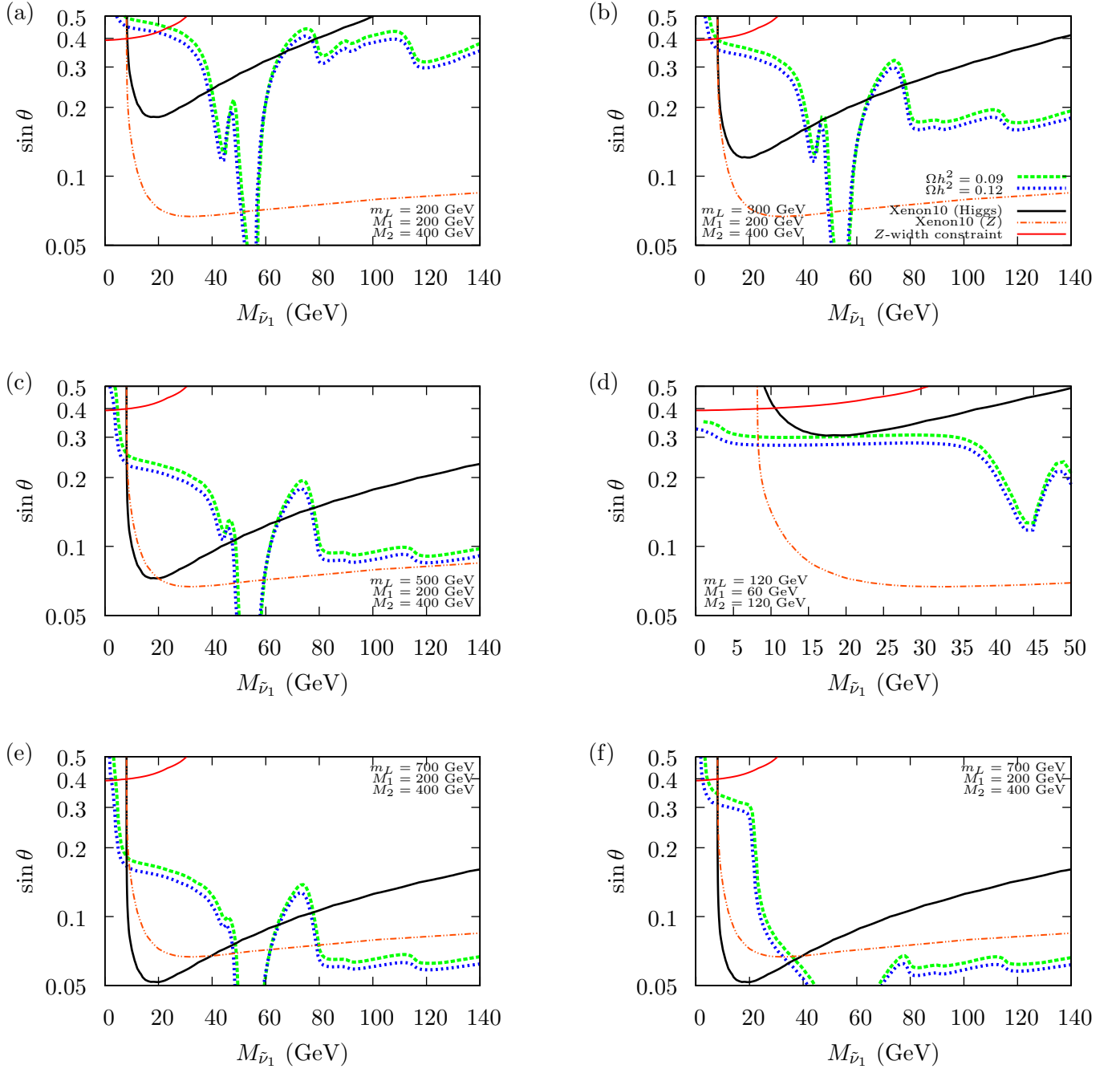
 
\centering
\begin{tabular}{ll}
\vspace{-.6 in} \hspace{0 in}(a) & (b)\\
\hspace{0 in}\input{figcosmoplot3} & \hspace{- 0 in} \input{figcosmoplot1} \\
\vspace{-.6 in} \hspace{0 in}(c) & (d)\\
\hspace{0  in}\input{figcosmoplot2} & \hspace{- 0 in} \input{figcosmoplot4} \\
\vspace{-.6 in} \hspace{0 in}(e) & (f)\\
\hspace{0  in}\input{figcosmoplot5} & \hspace{- 0 in} \input{figcosmoplotfathiggs}
\end{tabular}
\vspace{0 cm}
\caption{Constraints on the sneutrino parameter space from requiring the correct relic abundance, from direct-detection experiments, and from the invisible Z-width measurement.  The regions below the Z-width and direct-detection contours  are allowed.
The values taken for $m_L^2$, $M_1$, and $M_2$ are indicated in the plots, and the other MSSM parameters are as given in the text.  
}
\label{fig:abundance}
\end{figure}

In Fig.~\ref{fig:abundance} we display  regions in  $m_{\tilde \nu_1}$--$\sin \theta$ space that yield a relic abundance consistent with cosmological observations,
for various values of $M_1$, $M_2$, and $m_L^2$.  The other MSSM parameters are fixed as $\mu=300$ GeV, $\tan \beta = 10$, $m_A=500$ GeV,
$m_{l_R}=(300 \;\rm{GeV})^2$, $m_Q^2=m_{u_R}^2=m_{d_R}^2=M_3=(1 \;\rm{TeV})^2$, and $A_t =-1$ TeV, giving a Higgs mass of 116 GeV.  Also shown
are the constraints from the measurement of the invisible width of the $Z$ and from the recent results from the Xenon10 direct-detection experiment \cite{Angle:2007uj}.

Let us consider the plots of Fig.~\ref{fig:abundance}.
In plots (a)--(c), $M_1$ and $M_2$ are held fixed as three different values of $m_L$ are used.   Importantly, there are dramatic differences in the direct-detection constraints depending on whether elastic scattering via $Z$-exchange is suppressed -- in which case  Higgs exchange dominates --  or unsuppressed. We will discuss the circumstances in  which the $Z$-exchange contribution is suppressed in the next section.

If  the Higgs-exchange contribution dominates, we see in (a)--(c) that  $\tilde \nu_1$
masses   above the threshold for $W^+W^-$ production and near the $Z$ or Higgs poles are consistent with what we have learned about the dark matter abundance and with the latest Xenon10 results.  As $m_L$ is increased, the interesting regions in parameters space shift to smaller values of $\sin \theta$.
If instead  the scattering via $Z$-exchange is unsuppressed, only  the Higgs pole region is viable.

In (d) we focus on a particularly light spectrum (both for sneutrinos and gauginos).  In this case  one sees that without the $Z$-exchange contribution, sneutrinos with the appropriate relic abundance are allowed over the entire mass range. With unsuppressed $Z$-exchange scattering, one is forced into the light mass range ($m_{\tilde \nu} \lsim 10 \;\gev$). The precise mass below which this scenario is viable is not entirely certain, as the issue is sensitive to the highest velocity particles in the halo, for which a modified Gaussian is probably not a good description. 

In (e), we take $m_L$ to be very large. As a consequence, there is a large $A$-term for the same value of $\sin \theta$, making the $s$-channel Higgs annihilation more efficient, and allowing reasonable relic abundances for low values of $\sin \theta$. Consequently, a broad range of masses is viable, regardless of whether the scattering off of nuclei is dominated by $Z$- or Higgs- exchange.

Finally, in (f), we consider the effect of modifying the width of the Higgs. In various recent proposals \cite{Dermisek:2005ar,Dermisek:2005gg,Chang:2005ht,Schuster:2005py,Graham:2006tr,Chang:2006bw,Chang:2007de}, the Higgs width is dominated by final states other than $b \bar b$. This  possibility is motivated by the fine tuning problem associated with raising the Higgs mass above the LEP limit. For our purposes, the importance of non-standard Higgs decays  is that they can modify the form of the Higgs pole. We illustrate this by increasing the Yukawa coupling of the b-quark by a factor of five. One can see that this modification significantly impacts the allowed ranges for mixed-sneutrino dark matter -- comparing (e) and (f), the mass ranges from $40-50$ GeV and $60-80$ GeV open up. It is worth emphasizing that the uncertainties regarding the decays of the Higgs can have significant consequences for the allowed parameter space of any dark matter model which involves annihilation through an $s$-channel Higgs.

\begin{figure}[] 
\centering
\quad\quad\quad \quad\quad\quad \quad\quad\quad  \input{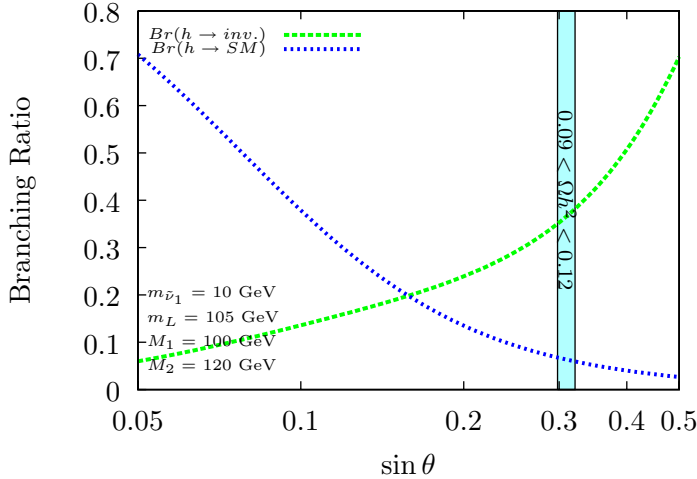}
\caption{Branching ratios for Higgs decays directly to standard model states, and to invisible final states via either  ${\tilde \nu}_1 {\tilde \nu}_1^*$ or ${\tilde \nu}_1 {\tilde \nu}_2$, for the parameters indicated.
}
\label{fig:higgsdecay}
\end{figure}

It is interesting to note that for the parameters used
for  Fig.~\ref{fig:abundance}(d), with $m_{\tilde \nu_1} \sim 10$~GeV and  $\sin \theta$ chosen to give the preferred dark matter abundance, the lightest Higgs boson decays invisibly, to ${\tilde \nu}_1 {\tilde \nu}_1^*$, more
than 80\% of the time.  If $m_L$ is lowered further, the Higgs also develops an appreciable branching ratio into ${\tilde \nu}_1 {\tilde \nu}_2$ final states.  Depending on
the spectrum -- for example, on whether decays to  $\chi^0_1$ are accessible -- ${\tilde \nu}_2$ may decay dominantly to ${\tilde \nu}_1 Z^{*}$.  In this case,  Higgs decays to ${\tilde \nu}_1 {\tilde \nu}_2$ would most often produce  a rather nondescript final state, although 20\% of the time the $Z^*$ would decay to neutrinos, giving an additional contribution to the invisible width of the Higgs.  In Fig.~\ref{fig:higgsdecay}, we take $Br({\tilde \nu}_2\rightarrow {\tilde \nu}_1 Z^{*}) = 100\%$, $m_L=105$~GeV, and $m_{{\tilde \nu}_1}=10$~GeV , and plot the branching ratios for a $116$ GeV Higgs to decay (i) invisibly and (ii) directly to standard model states, as functions of $\sin\theta$.  These branching ratios do not sum to unity because they exclude Higgs decays to  ${\tilde \nu}_1 {\tilde \nu}_2$ with ${\tilde \nu}_2$ decaying visibly.  In fact, we see that for the masses chosen for Fig.~\ref{fig:higgsdecay}, and for the values of  $\sin\theta$ preferred by cosmology,  this third class of decays dominates.

\subsection{Direct and Indirect Detection of Mixed Sneutrinos}
Cryogenic detectors such as CDMS and liquid noble gas detectors such as XENON have made considerable strides since mixed sneutrinos were originally considered.
 In this section we consider the cross section for $\snu_1$ -- nucleon scattering, paying particular attention to the assumptions built in. As indicated in the previous section, we find, consistent with \cite{Arina:2007tm}, that broad parameter regions remain viable for mixed-sneutrino dark matter, and that broader regions open up in the lepton-number violating case.

Including only the $Z$-exchange contribution, the cross section for $\snu_1$ to scatter off of nuclei is
\be
\sigma = \frac{G_F^2}{2 \pi} \mu^2 \left[ (A-Z)-(1-4 \sin^2 \theta_W) Z\right]^2 \sin^4 \theta,
\ee
where $\mu$ is the $\snu_1$ -- nucleus reduced mass.  As discussed above,
 this cross section exceeds  experimental limits for most parameters ranges, with exceptions at small $m_{\snu_1}$, near the Higgs pole, or with heavy left-handed sneutrinos.

Lepton number violation in the sneutrino mass matrix lifts mass degeneracy between the scalar and pseudo-scalar components of the lightest sneutrino, and scattering via $Z$-exchange  occurs inelastically (i.e., through a transition from the scalar to the pseudoscalar or vice-versa) \cite{Hall:1997ah,Smith:2001hy,TuckerSmith:2004jv}. As a consequence, particles with velocities below
\be
\beta_{min} = \sqrt{\frac{1}{2 M_N E_R}} \left( \frac{M_N E_R}{\mu} + \delta \right)
\ee
are incapable of scattering. Here, $M_N$ and $E_R$ are the target nucleus mass and recoil energy, and $\delta$ is the mass splitting between the scalar and pseudoscalar.  
Given that there are not expected to be any particles in the halo with velocities above the galactic escape velocity\footnote{Or, in our reference frame, $v_{esc}+v_{rot}$, where $v_{rot}$ is the net total velocity from motion of the Earth about the sun, and of the sun about the galactic center.}, by dialing $\delta$ large (of  order 100 keV), one can evade all direct detection constraints.

In this case, scattering from Higgs exchange still constrains the theory \cite{ArkaniHamed:2000bq}.
In the decoupling limit ($m_A \gg m_h$),
the  cross section  for Higgs-mediated $\tilde \nu_1$--nucleon scattering is
\begin{equation}
\sigma ={g_{h NN}^2\over 4 \pi} \left({g_{h \tilde{\nu}_1 \tilde{\nu}_1} \over m_N + m_{\tilde{ \nu}_1} }\right)^2 {m_N^2\over m_h^4},
\end{equation}
where $m_N$ is the nucleon mass, $g_{h NN}$ is the Higgs--nucleon coupling, and $g_{h \tilde{\nu}_1 \tilde{\nu}_1}$ is the coupling of the light Higgs boson to the LSP sneutrino,
\begin{equation}
g_{h \tilde{\nu}_1\tilde{\nu}_1} = -{m_Z^2 \over v }\cos 2\beta \sin^2\theta + {A \over  \sqrt{2}} \sin \beta \sin 2 \theta.
\end{equation}
The value of  $g_{h NN}$ is subject to
rather large uncertainties. Here we  adopt the up and down quark, strange quark, and heavy quark contributions to this coupling given in Refs.~\cite{Cheng:1988im}, \cite{Hatsuda:1990uw}, and \cite{Shifman:1978zn}, respectively.  This yields $g_{h NN}=1.26 \times 10^{-3}$.  Written in terms of this reference value, the cross section is
\begin{equation}
\sigma =\left({g_{h NN} \over 1.26 \times 10^{-3}} \right)^2 \left({g_{h \tilde{\nu}_1 \tilde{\nu}_1} \over m_N + m_{\tilde{ \nu}_1} }\right)^2 \left({115 \;{\rm GeV} \over m_h}\right)^4 (2.48\times 10^{-43} \; {\rm cm}^2).
\end{equation}

\subsubsection{Relation to neutrino mass}
In the lepton-number-violating case  gaugino loops generate  neutrino mases \cite{ArkaniHamed:2000kj,Borzumati:2000mc}. In the regime in which the diagram with a pure Wino running in the loop dominates, the correction to the neutrino mass is
\be
m_\nu = \frac{g^2 \sin^2 \theta \delta m_1 m_2}{32 \pi^2} \sum_{ij} f_{ij},
\label{eq:radnumass}
\ee
where
\be
f_{ij} = \frac{m_{\tilde W}^2 m_i^2 \log[m_{\tilde W}^2/m_i^2]+m_j^2 m_i^2 \log[m_i^2/m_j^2]+m_{\tilde W}^2 m_j^2 \log[m_j^2/m_{\tilde W}^2]}{(m_{\tilde W}^2 - m_i^2)(m_{\tilde W}^2 - m_j^2)(m_i^2-m_j^2)},
\ee
$m_{1,2}$ are the masses of the light and heavy complex eigenstates, and $\delta$ is the splitting between the scalar and pseudoscalar components of $\snu_1$.

For much of the parameter space, a splitting of order 100 keV is necessary to ensure that inelastic scattering at XENON is kinematically impossible. With the large mixings (and thus large $A$-terms) necessary to achieve the appropriate relic abundance, this mass splitting generates a neutrino mass of order$1\ \ev$. Combined with the limits on neutrino mass from cosmology \cite{Tegmark:2006az}, which are roughly $1\ \ev$ for the {\em sum} of the neutrino mass this possibility seems excluded. This has been emphasized recently by \cite{Arina:2007tm}. In particular, the authors of  \cite{Arina:2007tm} argue that inelasticity consistent with neutrino mass bounds  change the allowed parameter space very little. Here we review some important caveats to this.

One point  is that the standard value used for  galactic escape velocity, 650 km/s, may be too large. Indeed, the most recent simulations of Milky Way type galaxies \cite{Governato:2006cq} produce lower escape velocities ($\sim$ 450 km/s). Using the distributions of these simulations, rather than the much older halo parameters typically used  to set limits, the allowed parameter regions for mixed-sneutrino dark matter do expand. For example, the parameters $m_{\snu_1}=100\ \gev$, $\sin \theta =0.18$, $m_L=300\ \gev$, $M_2=400 \ \gev$, and $\delta = 50 \ \kev$ give a realistic relic abundance and a direct-detection rate  that is borderline at XENON, but only generate a neutrino mass of $0.32 \ \ev$. Although this mass pushes up against the cosmological limits, this example illustrates how the impact of $\delta$ on direct detection rates is highly sensitive to assumptions about the halo.

A second point is that the radiatively generated neutrino mass  is suppressed as the neutralino masses are increased, whereas the $\snu_1$ annihilation rate is insensitive to these masses if it is dominated by $s$-channel Higgs exchange.  For example, for the same parameters as in the previous paragraph, raising $M_2$ to 1 TeV does not change the relic abundance significantly, but does reduce the neutrino mass to 0.15 eV.  

Another possibility is that there might be an enhanced annihilation rate at smaller values of $\sin \theta$. 
For example, if the neutrino Yukawa couplings are large, as described in section \ref{sec:yuk}, there are additional contributions to the annihilation rate coming from the $\tilde \nu_1 \tilde \nu_1 h$ coupling $2\lambda^2 v \cos^2 \theta$. As this coupling can be parametrically comparable to $A \sin 2 \theta$, the annihilation via s-channel Higgs can be considerably enhanced, even at smaller mixing angles. As previously discussed, we leave the analysis of the model with large neutrino Yukawa couplings for future work.

Finally, the neutrino mass of equation~(\ref{eq:radnumass}) requires a Majorana mass insertion. If the gauginos are Dirac,  as described in \cite{Fox:2002bu},  a radiative mass will not be generated. In these scenarios, one can disregard the radiative neutrino mass entirely, even for large $\delta$.

When we consider the collider signatures of mixed sneutrinos, some of the parameter points we will study can accommodate
mixed-sneutrino dark matter only if the scattering via $Z$ exhange is strongly suppressed, possibly leading to a radiative neutrino mass that is too large.  Our main motivation for studying these parameter points is that LHC signatures for mixed sneutrinos are of interest in their own right, independent of the connection to dark matter.  However, we also believe that because of the myriad astrophysical and particle physics uncertainties, a liberal take on which regions of parameter space may be cosmologically interesting is warranted. 

\subsubsection{Indirect Constraints}
Indirect-detection experiments can also place important constraints on mixed-sneutrino dark matter; see  \cite{Arina:2007tm} for a thorough discussion. Here we focus on the scenario in which inelasticity is relevant.

If ${\tilde \nu}_1$ particles are captured by the sun at a large enough rate,
high-neutrinos produced in their decays can be detected on Earth.  We will allow for the possibility that capture
via $Z$ exchange is suppressed by the mass splitting between the scalar and pseudoscalar components of $\tilde{\nu}_1$,
so that the capture rate is determined by the contribution from Higgs exchange.  In this case we find, following \cite{Jungman:1995df}, that
the most stringent bounds from indirect detection experiments are not competitive with those from direct detection.  For example,
taking the parameters used for Figs.~\ref{fig:abundance}(b), with $m_{{\tilde \nu}_1}=100$~GeV and   $\sin\theta$ chosen to give the desired relic abundance,
we find a flux of upward through-going muons that is almost two orders of magnitude below the limits given in \cite{Desai:2004pq}.
With  the parameters used for Fig.~\ref{fig:abundance}(d), and taking $m_{{\tilde \nu}_1}=10$~GeV, the sneutrinos now annihilate directly to
neutrinos, but the predicted flux is still around an order of magnitude or more below current limits, depending on the flavors of neutrinos produced in
the annihilations.

As noted earlier, if the Higgs decays principally in non-standard fashion, then annihilation into neutrinos would be similarly suppressed if annihilation occurs through $s$-channel Higgs, although this certainly depends sensitively on the decay products of the Higgs boson. 

\subsection{LSP gravitinos}
 In supersymmetric theories in general, an intriguing possibility is that the true LSP is the gravitino, but that the lifetime of the NLSP is sufficiently long that the dark matter relic abundance is determined entirely by the freezeout of the NLSP \cite{Feng:2004zu,Feng:2004mt}. Typically, the NLSP is  imagined to be a stau or neutralino, but a mixed sneutrino could similarly serve as NLSP, decaying harmlessly into neutrino-gravitino.  The sneutrino NLSP case was considered within the MSSM in \cite{Covi:2007xj}. 
 
In  this gravitino-LSP scenario the parameter space  opens up dramatically, including regions with otherwise too large relic abundance, or regions where the XENON limits would have excluded mixed-sneutrino dark matter. This  gives us extra  motivation to be open-minded when studying the  collider phenomenology of mixed sneutrinos. Unfortunately, this gravitino LSP scenario leads to no signals at dark matter detectors, either direct or indirect.

\section{LHC signatures}
\label{sec:lhc}

What experimental signatures for mixed sneutrinos might be observed  these  at the LHC? This depends on the superpartner spectrum and, crucially, on the amount of mixing between the sterile and active sneutrinos.  In section  \ref{subsection:dilepton} we consider the case where the mixing angle $\theta$ is quite small. 
In this case the mostly-sterile sneutrinos will be
produced only rarely in cascade decays, unless they are the lightest superpartners. 
If they {\em are} the lightest superpartners, they will be produced in the decays of the NLSP,
and the collider signatures depend on the identity of that particle. 
In section \ref{subsection:dilepton} we will see that if the NLSP is  a right-handed slepton,  a distinctive opposite-sign dilepton signature potentially emerges.
If the NLSP is instead  $\chi^0_1$, the collider phenomenology will be  the same as with a neutralino LSP, but even in this case, there is a simple point to be made:  a given cosmologically  disfavored point in MSSM parameters space may become cosmologically viable with a mixed-sneutrino added at the bottom of the spectrum.    

In sections \ref{subsection:jetlepton} and \ref{subsection:zh}, we consider additional signatures that become possible if the mixing angle is larger, $\theta \gtrsim 0.1$.   We have seen that, for these larger mixing angles, there is tension between having the correct relic abundance, evading direct-detection experiments, and satisfying the
neutrino-mass bound.  While in certain variations of the model this tension
may be eliminated ({\em e.g.} in a scenario with Dirac gauginos),  we regard these mixed-sneutrino signatures as important to study independent of whether the sneutrinos produced are the cold dark matter, for the following reasons:
\begin{itemize}
\item   Discovering mixed-sneutrinos at the LHC  would shed important light on the nature of neutrino masses; it would suggest that  the neutrinos are of Dirac type, or else, that the seesaw scale is not be much larger than the weak scale.  
\item Even if the sneutrinos produced at the LHC are not the dark matter, they could still be relevant to the dark matter question.  Here are two possible scenarios that illustrate this point: (1)  The lightest mixed-sneutrino is the NLSP.  It freezes out in the early universe and then decays to gravitino dark matter.   (2) In addition to  the sneutrinos with large enough mixing angles to be produced
at the LHC, there is a lighter one with a smaller active component, suitable to be the cold  dark matter. 
\end{itemize}
In section  \ref{subsection:jetlepton}  we consider kinematic edges that can appear in jet-lepton invariant mass distributions when charginos decay directly to mixed sneutrinos.  We also discuss the possibility of using mass estimation techniques to distinguish this signature from MSSM ones.  In section  \ref{subsection:zh}  we study signatures associated with  decays
of the heavier sneutrinos to the lighter ones, involving Higgs and $Z$ bosons .

\subsection{Dilepton mass distributions}
\label{subsection:dilepton}
Assuming that a mixed-sneutrino is the LSP, its presence at the end of every cascade decay chain makes for SUSY signals rich in leptons.  In this section, we  consider cascades involving a right-handed slepton NSLP, 
and find the following:
\begin{itemize}
\item{Opposite-sign same-flavor (OSSF) dileptons can arise from a two-body decay followed by a three-body decay, whereas in the MSSM they typically come from  two-body followed by two-body, or from  a single three-body decay. If the final state leptons are mostly $\mu$ or $e$, this allows us to distinguish the mixed-sneutrino scenario from the MSSM over significant regions of parameter space.}
\item{If ${\tilde \tau}_1$ is produced, rather than ${\tilde e}_R$ or ${\tilde \mu}_R$,   a prominent signal is still possible and may be distinguishable from the MSSM.}
\end{itemize}

If the LSP is a weakly mixed sneutrino and the NLSP is a right-handed slepton, we have the following
possible decays for the lightest neutralino: 
\begin{equation}
\chi^0_1 \rightarrow {\tilde \nu}_1 \nu  \quad\quad    \chi^0_1 \rightarrow  {\tilde \tau}_1 \tau                                 \quad\quad  \chi^0_1 \rightarrow  {\tilde l}_R  l,
\end{equation} 
where $l=e,\mu$.  The direct decay to  ${\tilde \nu}_1 \nu$  is suppressed by the small mixing angle, leaving ${\tilde \tau}_1 \tau$ and ${\tilde l}_R  l$ as the competing decay channels.  We will assume for
now that $ \chi^0_1 \rightarrow  {\tilde l}_R  l$ has a substantial branching ratio, and consider  the case
where $\chi^0_1 \rightarrow  {\tilde \tau}_1 \tau $ completely dominates at the end of this section.

Under this assumption, we expect  a large number of right-handed sleptons to be produced at the LHC.  How do they decay?
The possibilities are
\begin{equation}
{\tilde l}_R \rightarrow {\tilde \nu}_1 W  \quad\quad   {\tilde l}_R \rightarrow  {\tilde \tau}_1 \tau  l                               \quad\quad {\tilde l}_R \rightarrow  {\tilde \nu}_1 \nu  l.
\end{equation} 
Even if the two-body decay to ${\tilde \nu}_1 W$ is kinematically accessible, it is not only suppressed by the sneutrino mixing angle, but also vanishes in the absence of left-right slepton mixing.  For $\theta \sim 0.05$, it is typically negligible compared to the three-body decays for $l=e$, and only potentially competitive for $l= \mu$.  Note also that the relevant coupling here depends on flavor issues -- if the active component of ${\tilde \nu}_1$ is entirely third-generation, these decays are absent.  The second decay, to ${\tilde \tau}_1 \tau  l $, may or may not be kinematically allowed.  Even if it is allowed, its kinematical suppression can easily make the third decay, to ${\tilde \nu}_1 \nu  l$, the dominant one.
As we will illustrate by example below, this is true even though the decay to  ${\tilde \nu}_1 \nu  l$
is mixing-suppressed.  

Assuming, then, that $\chi^0_1 \rightarrow  {\tilde l}_R  l$ and  ${\tilde l}_R \rightarrow  {\tilde \nu}_1 \nu  l$ both have substantial branching ratios,  an opposite-sign same-flavor (OSSF) dilepton signature results.  
The $l^+l^-$ invariant mass distribution is predicted to have  a kinematic endpoint at
\begin{equation}
m_{l^+l^-}^{max} = m_{\chi_1^0} \sqrt{1 - (m_{\tilde l_R}/m_{\chi_1^0})^2} \sqrt{1-(m_{\tilde \nu_1}/m_{\tilde l_R})^2}.
\end{equation}
In supersymmetric models, OSSF dilepton signatures, with associated kinematic edges, are quite common.
How distinctive is the dilepton mass distribution in the case with a mixed-sneutrino LSP?

In standard SUSY models, an OSSF dilepton signature can arise via the sequence of two-body
decays  $\chi_2^0 \rightarrow (\tilde l)  l^+ \rightarrow (\chi_1^0  l^-) l^+ $.  One important difference compared to this standard case is simply that in the mixed-sneutrino LSP case, the dileptons come from $\chi_1^0$ decays.  Provided the relevant branching ratios are sizable
we would thus typically expect a larger lepton multiplicity than in the case where $\chi_2^0$ initiates the decays.  However, $\chi_1^0$ decays  can also produce an OSSF dilepton signature in a scenario  with a gravitino LSP and a right-handed slepton NLSP, through the sequence $\neutone\rightarrow  ({\tilde l}_R ) l^+ \rightarrow  ({\tilde G}  l^-) l^+$ (prompt decays of the NLSP slepton are possible for a low SUSY-breaking scale). 

\begin{figure}[htbp] 
   \centering
   \includegraphics[width=2in]{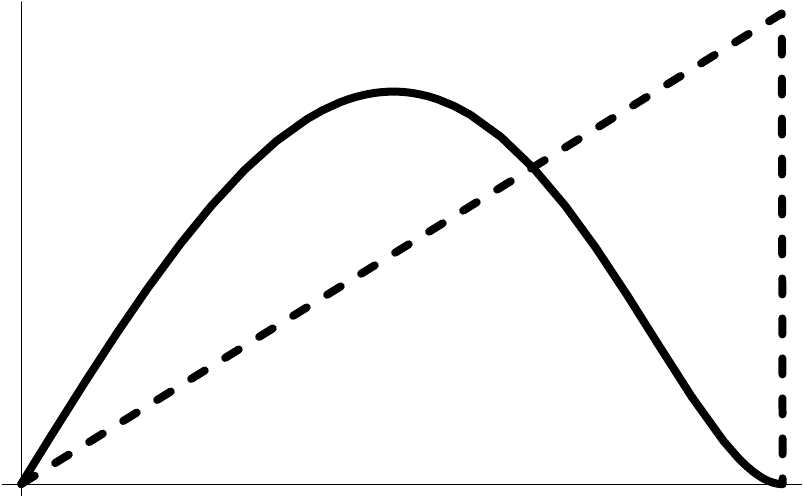} 
   \caption{OSSF dilepton invariant-mass distributions from the sequence of two-body decays
   $\chi_2^0 \rightarrow( \tilde l)  l^+ \rightarrow( \chi_1^0  l^-) l^+ $ (dashed), and from the two-body/three-body sequence $\chi^0_1 \rightarrow  ({\tilde l}_R)  l^+ \rightarrow ( {\tilde \nu}_1 \nu  l^-) l^+ $ (solid).
   For the latter, the amplitude of the three-body decay is set to be constant, and we take $m_\slepr=0.8 \;m_\neutone$ and $m_\snu=0.5\; m_\neutone$. 
    }
   \label{fig:22-23}
\end{figure}

These scenarios are easily distinguished from the mixed-sneutrino case by their dilepton invariant-mass distributions.
The  two-body/two-body sequences  $\chi_2^0 \rightarrow( \tilde l)  l^+ \rightarrow (\chi_1^0  l^-) l^+ $ and
$\neutone\rightarrow ( {\tilde l}_R ) l^+ \rightarrow ( {\tilde G}  l^-) l^+$ both have the distribution 
\begin{equation}
{dP \over d m_{l^+ l^-}} \propto  m_{l^+ l^-}.
\end{equation}
If we take the matrix element of the three-body decay to be constant and just consider the phase-space dependence, the   two-body/three-body sequence $\chi^0_1 \rightarrow ( {\tilde l}_R ) l^+ \rightarrow ( {\tilde \nu}_1 \nu  l^-) l^+ $ gives
\begin{equation}
{dP \over d x} \propto  x  
\left(1-x^2- \mu^2\left[1+  \ln\left(\frac{1-x^2} {\mu^2}\right)  \right] \right),
\label{eq:twothree}
\end{equation} 
 where we have defined $x=m_{l^+ l^-}/\sqrt{m_\neutone^2-m_\slepr^2}$,  $\nu = m_\slepr/m_\neutone $, and  $\mu = m_\snu/m_\slepr$. 
 Normalized plots of these very different looking  distributions are shown in Figure \ref{fig:22-23}.  
 
\begin{figure}[htbp] 
   \centering
    \includegraphics[width=3in]{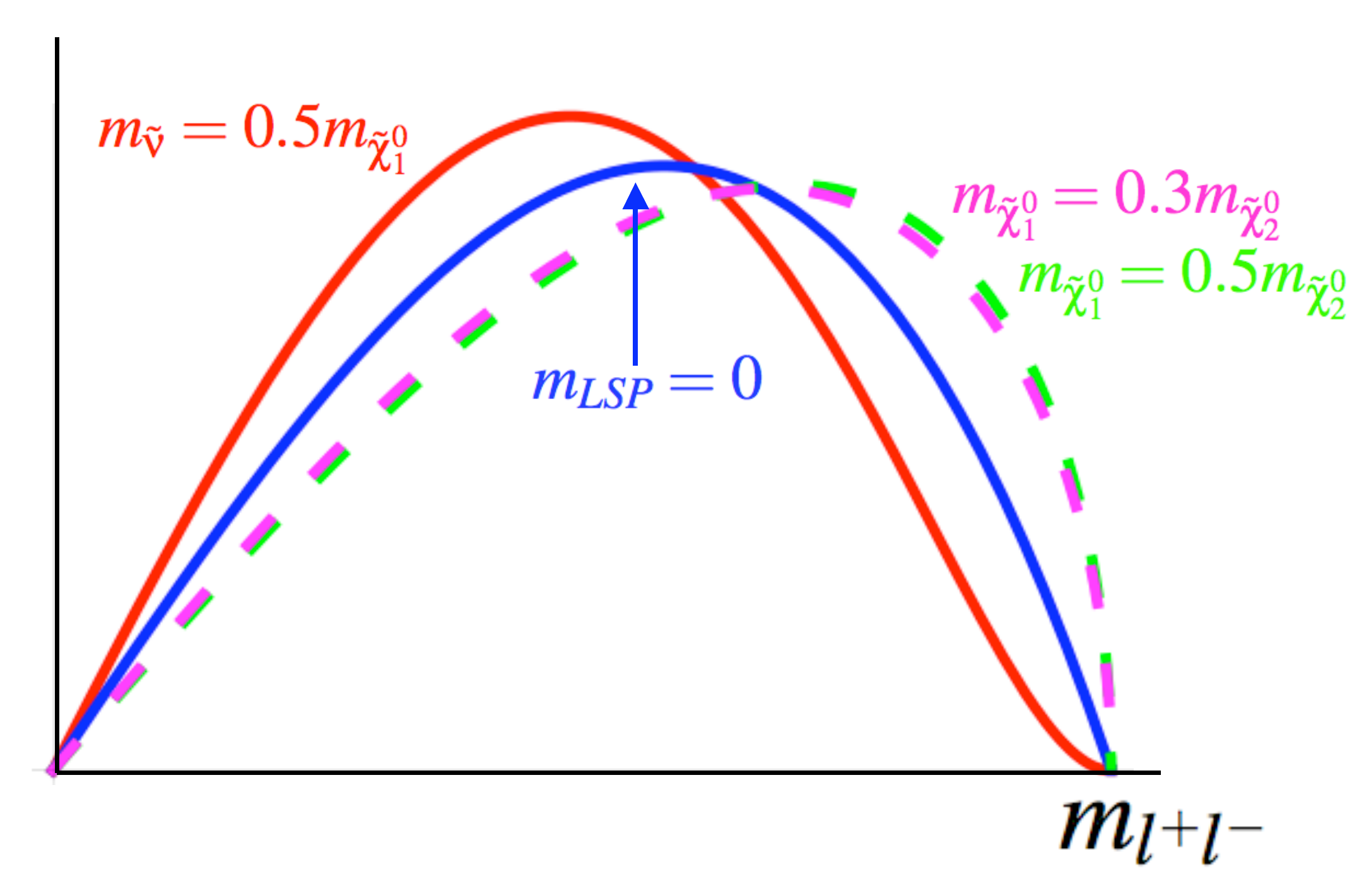}
   \caption{OSSF dilepton invariant-mass distributions for the three-body decay
   $\chi_2^0 \rightarrow \chi_1^0  l^+ l^- $ (dashed), and for the two-body/three-body sequence $\chi^0_1 \rightarrow  ({\tilde l}_R)  l^+ \rightarrow ( {\tilde \nu}_1 \nu  l^-) l^+ $ (solid).
   For the latter, the amplitude of the three-body decay is set to be constant, and we take $m_\slepr=0.8 \;m_\neutone$. The distributions are identical for the massless-LSP case, shown in the middle.}
   \label{fig:3-23compare}
\end{figure}

A softer dilepton invariant-mass distribution arises in the MSSM if the leptons come from a three-body decay such as  $\chi_2^0  \rightarrow \chi_1^0  l^+ l^- $. In this case the endpoint of the distribution is just the mass difference between two neutralinos,
\begin{equation}
m_{l^+l^-}^{max} = m_{\chi_2^0} -m_{\chi_1^0}.
\end{equation}
Again taking the matrix element of the three-body decay to be constant and considering the phase-space dependence alone, the dilepton invariant-mass distribution is
\begin{equation}
{dP \over d x} \propto  x \sqrt{\left( 1-x^2\right) \left( (1-K^2)^2-x^2 \right)}, 
\end{equation}
where  $x=m_{l^+ l^-}/(m_\neuttwo-m_\neutone)$ and $K=2m_\neutone/(m_\neuttwo-m_\neutone)$.  
In the massless-LSP limit, this distribution and the two-body/three-body distribution of equation
(\ref{eq:twothree}) both  reduce to 
\begin{equation}
{dP \over d x} \propto  x ( 1-x^2),
 \end{equation}
 where $x=m_{l^+ l^-}/m_{l^+l^-}^{max}$.
Although these distributions are identical in the massless-LSP limit, they shift in opposite directions
as the LSP mass increases, as shown in figure~\ref{fig:3-23compare}.  
There we see that the mixed-sneutrino distributions are significantly softer than what the MSSM three-body decays give, unless the LSP neutralino mass for the MSSM case is unusually small.  Other  observables, such as the overall lepton multiplicity, would likely help to further distinguish particular points in the parameter spaces of these two scenarios.  Moreover, as we will see,  the observed value of the kinematic endpoint of the $m_{l^+ l^-}$ distribution may not be easily reconciled with a very small   value of $m_\neutone/m_\neuttwo$.

We have used Monte Carlo simulation to explore the distinguishability of these scenarios further.  We modified the SUSY-HIT package \cite{Djouadi:2006bz} for calculating superpartner masses and decay branching ratios  to incorporate mixed sneutrinos and the associated decays.  For example,
the three-body decays of right-handed sleptons were implemented by appropriately modifying the the matrix elements given for three-body squark decays.
We generated events with full SUSY production using Pythia 6.4 \cite{Sjostrand:2006za}, with the pytbdy.f file modified to include the momentum dependence in the 
amplitudes for  the three-body  decays of right-handed sleptons via off-shell bino, 
\begin{eqnarray}
|\mathcal{M}(\slepr \rightarrow \snu {\overline \nu} l)|^2 & =  & 4 {g'}^4 m_\slepr^2 \; \frac{4 E_l E_\nu - m_{l\nu }^2}{\left(m_\neutone^2 - m_{\nu \snu}^2\right)^2} \\
|\mathcal{M}(\slepr \rightarrow \snu^*\nu l)|^2 &  = &  4 {g'}^4\;  \frac{m_\neutone^2 \; m_{l \nu}^2}{\left(m_\neutone^2 - m_{\nu \snu}^2\right)^2}. 
\end{eqnarray}
Here  $m_{l\nu }$ and  $m_{\nu \snu}$  are the lepton-neutrino and neutrino-sneutrino invariant masses, respectively.
The momentum dependence for three-body neutralino decay amplitudes is already included in Pythia.   After generating the fully showered and hadronized Pythia events, we passed these to the PGS 4.0 detector simulator \cite{PGS}, taking the granularity of the calorimeter grid to be $\delta \eta \times \delta \phi = 0.1\times 0.1$.

For the mixed-sneutrino case, we start with the mSUGRA-like high-scale parameters ${\tilde m}^2=(10$ GeV$)^2$, $M_{1/2}=450$ GeV, $\tan\beta =10$, $A_t=-500$ GeV, and
$A_b=A_\tau=0$. Then we add a weakly mixed sneutrino with $\theta=0.05$ and $m_{\tilde \nu_1}=58$ GeV to the bottom of the resulting spectrum.  The physical superpartner masses for this point are given in Table \ref{tab:spectrum}.
\begin{table}[htdp]
{\begin{tabular}{||c||c||}
\hline
\hline
$m_{\tilde g}$ & 1039 \\ 
\hline
$m_{{\tilde \chi}_2^\pm}$ & 678 \\
$m_{{\tilde \chi}_1^\pm}$ & 349 \\ 
\hline
$m_{{\tilde \chi}_4^0}$ & 678 \\
$m_{{\tilde \chi}_3^0}$ & 668 \\ 
$m_{{\tilde \chi}_2^0}$ & 350 \\
$m_{{\tilde \chi}_1^0}$ & 184 \\
\hline
\hline
\end{tabular} 
\hspace{0.3in}
\begin{tabular}{||c||c||}
\hline
\hline
$m_{{\tilde u}_L}$ & 948 \\ 
$m_{{\tilde u}_R}$ & 915 \\
\hline
$m_{{\tilde d}_L}$ & 952 \\ 
$m_{{\tilde d}_R}$ & 912 \\
\hline
$m_{{\tilde t}_2}$ &914 \\ 
$m_{{\tilde t}_1}$ & 663 \\
\hline
$m_{{\tilde b}_2}$ &910 \\ 
$m_{{\tilde b}_1}$ & 860 \\
\hline
\hline
\end{tabular}
\hspace{0.3in}
\begin{tabular}{||c||c||}
\hline
\hline
$m_{{\tilde l}_L}$ & 303 \\ 
$m_{{\tilde l}_R}$ & 172 \\
\hline
$m_{{\tilde \tau}_2}$ & 306 \\ 
$m_{{\tilde \tau}_1}$ & 162 \\
\hline
$m_{\snu_2}$ &293 \\ 
$m_{\snu_1}$ & 58 \\
\hline
\hline
\end{tabular}
\hspace{0.3in}
\begin{tabular}{||c||c||}
\hline
\hline
$m_{H^\pm}$ & 730 \\ 
$m_{H}$ & 726 \\ 
$m_{A}$ & 726 \\ 
$m_{h}$ & 116 \\ 
\hline
\hline
\end{tabular}}
\caption{Superpartner and Higgs boson masses for the parameter point used to study the dilepton signature in the mixed-sneutrino case. All masses
are in GeV.}
\label{tab:spectrum}
\end{table}%

The most important masses for the dilepton signature are $m_\neutone$, $m_\slepr$, and $m_{\snu_1}$, whose values lead to a kinematic endpoint
of $m_{l^+l^-}^{max} =63$ GeV.    For the parameter point chosen, the branching ratios for the lightest neutralino are $Br(\chi^0_1 \rightarrow  {\tilde l}_R  l)= 41\%$,
$Br(\chi^0_1 \rightarrow  {\tilde \tau}_1 \tau) = 57\%$, and $Br(\chi^0_1 \rightarrow {\tilde \nu}_1 \nu )=2\%$.  The branching ratios for the right-handed selectrons are 
are $Br({\tilde e}_R \rightarrow  {\snu}_1 \nu e) = 97\%$ and  $Br({\tilde e}_R \rightarrow  {\tilde \tau}_1  \tau \nu) = 3\%$, and the branching ratios for the right-handed smuons are  $Br({\tilde \mu}_R \rightarrow  {\snu}_1 \nu \mu) = 69\%$,  $Br({\tilde \mu}_R \rightarrow  {\snu}_1 W) = 29\%$, and  $Br({\tilde \mu}_R \rightarrow  {\tilde \tau}_1  \tau \nu) = 2\%$.  The total widths for $\tilde{e}_R$ and $\tilde{\mu}_R$ are both hundreds of eV, so we can assume that they decay promply.
Note that although we mix in three generations of sterile sneutrinos, all with the same mixing angle,  the detectability of the dilepton signature does
not rely on this simplifying assumption.  For example, if there is only a single sterile sneutrino which mixes with the stau sneutrino alone, the branching ratios for ${\tilde \mu}_R \rightarrow  {\snu}_1 \nu \mu$ and ${\tilde e}_R \rightarrow  {\snu}_1 \nu e$ are both above 90\% -- the branching ratio actually goes up for $\tilde{\mu}_R$ because it can no longer go to ${\snu}_1 W$.

For this parameter point, we find using Prospino 2.0 \cite{prospino} that the NLO cross section for squark and gluino production \cite{Beenakker:1996ch,Beenakker:1997ut}  is 2.7 pb.  We generate 80,000 events, corresponding to $\sim 30$ fb$^{-1}$.  
In our analysis, we demand either an $e^+ e^-$ pair or a $\mu^+ \mu^-$ pair, where the leptons are required
to have $p_T>10$ GeV and $|\eta|<2.4$.  To suppress standard-model background,
we further require $\sum p_T>1500$ GeV, where  the sum is over jets with $p_T>20$ GeV,
leptons and photons with $p_T>10$ GeV, and missing $p_T$.  After this cut the leading
standard-model background is from $t{\overline t}$, which we also simulate using Pythia.
We generate $18.9 \times 10^6$ $t{\overline t}$  events, corresponding to $~23$ fb$^{-1}$ of integrated
luminosity, where  we take $\sigma=830$ pb for the $t{\overline t}$ production cross section at NLO
\cite{Bonciani:1998vc}.  After the cuts, we have 850 $t{\overline t}$ events after rescaling to  30 fb$^{-1}$, compared with
8,812 events from SUSY production.  

\begin{figure}[]
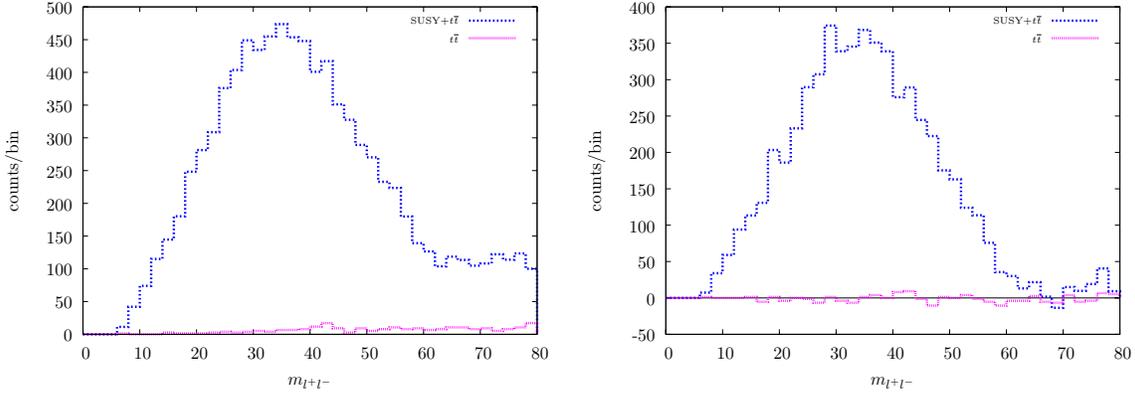
 
\centering
\scalebox{0.6}{\input{figsigbackshort}}
\scalebox{0.6}{\input{figsigbackshortsubtract}}
\caption{Left: OSSF dilepton invariant-mass distributions for SUSY events and ${t \overline t}$ background.  Right: flavor-subtracted opposite-sign dilepton invariant-mass distributions.}
\label{fig:dilepshort}
\end{figure}

In the first plot of  figure \ref{fig:dilepshort}, we show the OSSF dilepton invariant-mass distribution for events passing the cuts.
The distribution rapidly decreases as the mass approaches the expected endpoint $m_{l^+l^-}^{max} =63$ GeV
from below.  It then levels off due to the relatively large SUSY background.  This background can be dealt with
using a standard flavor subtraction, as shown in the second plot of the same figure.
To make that plot,
the analysis is redone, this time demanding either an $e^+ \mu^-$ pair or a $\mu^+ e^-$ pair.  The opposite-sign
dilepton invariant mass distributions for these events are then subtracted from the OSSF distribution.  We see that this procedure does a good job
reducing the SUSY background, and the kinematic endpoint is evident, within a few GeV or so of the expected
position at 63 GeV.

A simple modification to this parameter point is to remove  the mixed-sneutrino LSP with a light gravitino, 
with the assumption that ${\tilde e}_R$ and  ${\tilde \mu}_R$ decay promptly to it (this will not be the case if decays to ${\tilde \tau}_1$ are accessible).   
\begin{figure}[]
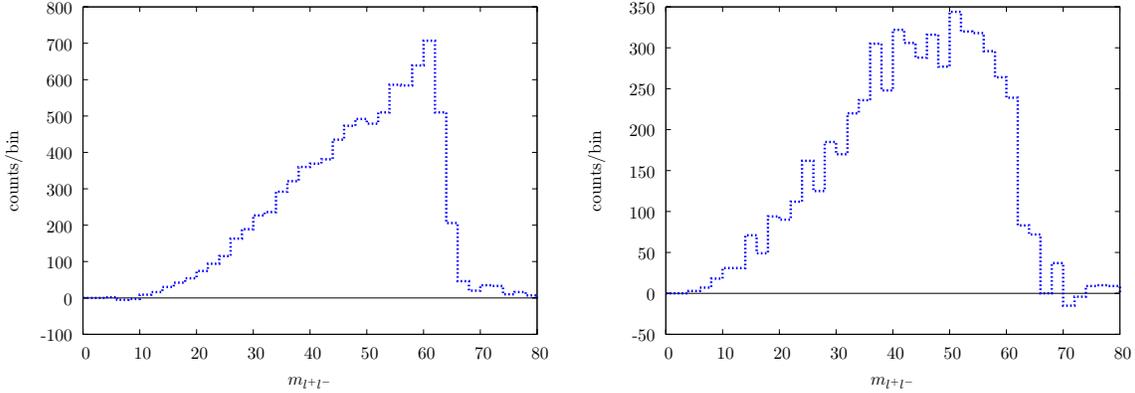
 
\centering
\scalebox{0.6}{\input{figgravdilep}}
\scalebox{0.6}{\input{figchi3boddilep}}
\caption{Flavor-subtracted OSSF dilepton invariant-mass distributions in (left) the gravitino LSP case and (right) the MSSM case with the
three-body decay $\chi_2^0  \rightarrow \chi_1^0  l^+ l^- $.}
\label{fig:gravdilep}
\end{figure}
If we adjust the right-handed slepton mass to keep the OSSF dilepton endpoint near 63 GeV,
we find the distribution shown in the first plot of figure \ref{fig:gravdilep}.
As expected, the differences compared to the the mixed-sneutrino case are clear due to the two-body kinematics of the relevant decays.

We now want to compare the mixed-sneutrino distribution with that from an MSSM parameter point in 
which the  three-body decays  $\chi_2^0  \rightarrow \chi_1^0  l^+ l^- $ are important.  So, we choose
low-scale parameters such that the most relevant physical masses are $m_{{\tilde \chi}_1^\pm}=101$  $m_\neuttwo=101$ GeV, $m_\neutone=37$ GeV, $m_{{\tilde l}_L}=129$ GeV,
and $m_{\tilde \nu}=102$ GeV.  The parameters are chosen so that the  splitting between $m_\neuttwo$ and $m_\neutone$ gives a kinematic endpoint very close
to the 63 GeV value from the mixed-sneutrino case.  Also, the gaugino masses are taken as small as  they can be consistent with
negative results from direct SUSY searches at LEP.  The rationale for doing this is to soften the dilepton invariant-mass distribution as  much as possible,
to see how similar to the mixed-sneturino distribution it can look (recall that the distributions should be very similar in the massless-LSP limit).   Another 
way to soften the $m_{l^+l^-}$ distribution is to make the intermediate slepton in the three-body decay just barely off-shell.  So, we make $m_L$ small; 
however, we want to ensure that the two-body decay $\chi_2^0 \rightarrow \tilde{ \nu} \nu$ is not kinematically accessible, which prevents us from lowering
$m_{{\tilde l}_L}$ arbitrarily close to $m_\neuttwo$.  Note that for the parameters chosen $m_\snu$ is just above $m_\neuttwo$.

The flavor-subtracted $m_{l^+ l^-}$ distribution for this set of parameters is shown in the second plot of figure \ref{fig:gravdilep}.
The distribution is softer than in the gravitino LSP case, but still easily distinguished from the mixed-sneutrino distribution of figure \ref{fig:dilepshort}.
Had the value of the kinematic endpoint for the mixed-sneutrino been significantly larger it would presumably allow one to find a point in the MSSM parameter
space that gives a more similar looking distribution.  For this reason we stress that other observables, such as the overall lepton multiplicity, may also be useful for 
distinguishing particular points in the two parameter spaces.  A detailed study of a number of relevant observables at once would likely
be efficient at ruling out candidate parameter points.  

For example, for the particular point we have chosen on the mixed-sneutrino side, an
 additional distinctive feature evident in the $m_{l^+l^-}$ distribution is a second kinematic
 edge at $\sim 140$ GeV.  
 \begin{figure}[] 
\centering
\scalebox{0.6}{\input{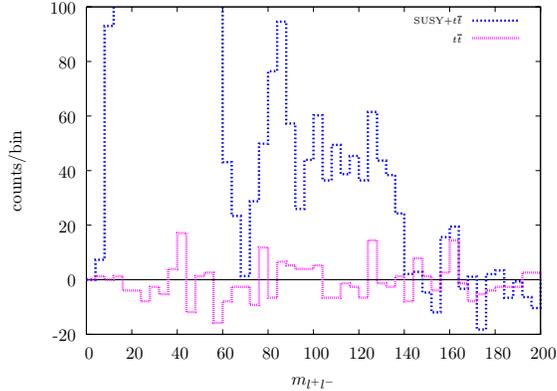}}
\caption{Flavor-subtracted OSSF dilepton invariant-mass distributions for SUSY events and ${t \overline t}$ background, in
the mixed-sneutrino case. A second kinematic endpoint is visible at $\sim 140$ GeV.}
\label{fig:secondedge}
\end{figure}
 This is shown in figure \ref{fig:secondedge}, which is essentially a zoomed-in 
 view of figure  \ref{fig:dilepshort}, going out to larger invariant masses.
 The flavor-subtracted distribution doesn't average to zero beyond $\sim 63$ GeV because
 OSSF dileptons can also come from the two-body/two-body sequence  
 $\chi_2^0 \rightarrow (\tilde l)  l^+ \rightarrow (\chi_1^0  l^-) l^+ $, and the kinematic endpoint for this 
 sequence is indeed near 140 GeV for the parameters chosen.    The possible presence of two edges,
 one associated with $\neuttwo$ decays and the other associated $\neutone$ decays, is one
 more puzzle piece one could use to distinguish points in mixed-sneutrino parameter space from 
 points in  MSSM parameter space.
 
 \vskip 0.15in
 \noindent {\it Decays to $\tau$'s}
 
 We conclude this section by considering the possibility that $ \chi^0_1 \rightarrow  {\tilde \tau}_1 \tau $
is the dominant decay of the lightest neutralino, with nearly 100\% branching ratio.   In this case, the
signature involving $e^+ e^-$ or $\mu^+ \mu^-$ pairs  is absent, and the situation becomes more challenging
experimentally.    If $\tilde{\tau}_1$ decays dominantly to $\snu_1 \nu \tau$, then one could hope to observe
an endpoint in the ditau invariant mass distribution.  However, if $\tilde{\tau}_1 \rightarrow \snu_1 W$ is kinematically
accessible, it is likely to dominate.  Here we assume this two-body decay is kinematically accessible and consider
the detectability of $\neutone \rightarrow ({\tilde \tau}_1) \tau^+   \rightarrow (\snu_1 W^-)\tau^+$, with the $W$ decaying leptonically.

To study this issue we take the high-scale parameters ${\tilde m}^2=(50$ GeV$)^2$, $M_{1/2}=350$ GeV, $\tan\beta =10$, $A_t=-500$ GeV, and
$A_b=A_\tau=0$. Then we add a weakly mixed sneutrino with $\theta=0.05$ and $m_{\tilde \nu_1}=51$ GeV.   For the
purposes of this study, the most important masses are  $m_\neutone=141$ GeV, $m_{{\tilde \tau}_1}=134$ GeV,
and $m_{{\tilde \nu}_1}=51$ GeV, and the most important branching ratios are $Br(\chi^0_1 \rightarrow  {\tilde \tau}_1 \tau)=94\%$ and $Br(\tilde{\tau}_1 \rightarrow \snu_1 W)=94\%$.  In the sequence $\chi^0_1 \rightarrow  ({\tilde \tau}_1) \tau^+ \rightarrow (\snu_1 [W^- ]) \tau^+ \rightarrow (\snu_1 [ l {\overline \nu} ] )\tau^+$, the invariant-mass of the final-state lepton and tau has an upper bound of 37 GeV for the masses considered.

\begin{figure}[]
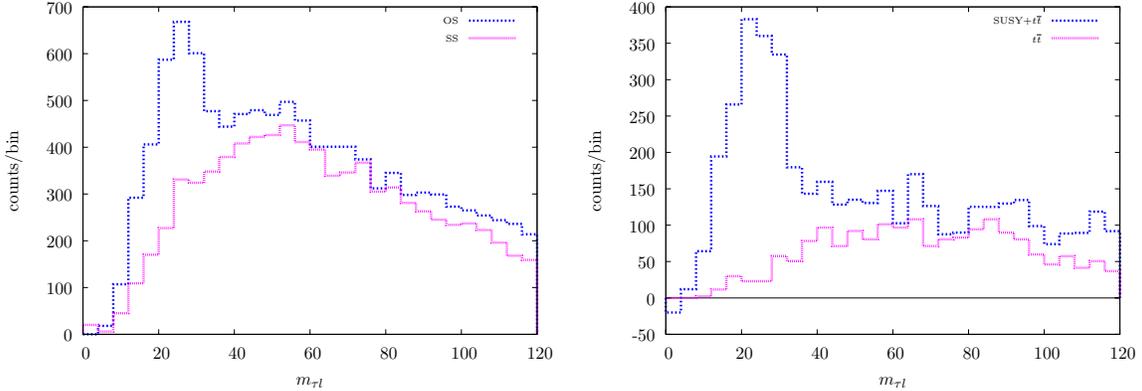
 
\centering
\scalebox{0.6}{\input{figtauWsigalone}}
\scalebox{0.6}{\input{figtauWsigback}}
\caption{Left: opposite-sign (OS) and same-sign (SS) lepton-tau invariant-mass distributions from SUSY production alone.
Right: opposite-sign minus same-sign distributions for SUSY plus $t {\overline t}$ production, and for $t {\overline t}$ production alone.}
\label{fig:tauWsigalone}
\end{figure}

We find the total SUSY production cross section for this point to be 10.6 pb for $pp$ collisions
at $\sqrt{s}=14$ TeV, 
and work with $\sim$ 587,000 events, corresponding to about 55 fb$^{-1}$ of integrated luminosity.
We take all events with at least one reconstructed $\tau$ and at least
one isolated lepton ($e$  or $\mu$),  and as before we put apply a cut $\sum p_T>1500$ GeV to reduce standard-model backgrounds.  The invariant mass $m_{\tau l}$  is calculated for all $\tau$-lepton pairings with opposite sign, giving the OS $m_{\tau l}$ distribution shown in the first plot of figure \ref{fig:tauWsigalone}.   Repeating the same procedure,  this time requiring the $\tau$ and the lepton to have the {\em same} sign, gives the SS distribution in the same plot.   The peak at low invariant-mass in the OS distribution
arises from events with $\chi^0_1 \rightarrow  ({\tilde \tau}_1) \tau^+ \rightarrow (\snu_1 [W^-])  \tau^+ \rightarrow ( \snu_1[ l {\overline \nu} ]) \tau^+$.  The OS distribution is similar to the SS distribution beyond
this peak, but an excess in the OS distribution over the SS distribution does persist beyond the
expected $37$ GeV endpoint, because there are other ways to produce opposite-sign leptons and taus in association with each other ($\chi^0_2 \rightarrow ( {\tilde \tau}_1) \tau^+ \rightarrow( \snu_1 [W^-])  \tau^+ \rightarrow (\snu_1 [l {\overline \nu}])  \tau^+$ being just one example). 

To gauge whether this signature would be observable above background, we include the $t {\overline t}$ 
sample generated with Pythia.   Taking the difference between the OS and SS distributions for both
the SUSY and $t {\overline t}$ backgrounds, we find the subtracted distributions shown in the second plot of figure \ref{fig:tauWsigalone}.  Despite the significant $t {\overline t}$ and SUSY backgrounds, a rather dramatic
fall-off in the distribution is evident for invariant masses between 30 and 40 GeV.  

In the MSSM, the decay $\snu_\tau \rightarrow {\tilde \tau}_1 W^+$ may occur due to the left-right stau mixing.   If the stau subsequently decays
as $\tau_1 \rightarrow \neutone \tau$, then we have associated $W$-$\tau$ production, just as
we had in the mixed-sneutrino case.  Here we do not explore in detail the extent to which this MSSM decay sequence could be distinguished from the mixed-sneutrino decay sequence considered above, but simply note that in the mixed-sneutrino scenario the  signal has the potential to be much more prominent, given that it originates from $\neutone$
decays rather than $\snu_\tau$ decays.

\subsection{Jet-lepton mass distributions}
\label{subsection:jetlepton}
At the LHC, most SUSY evens will begin with squark or gluino production, and so leptons produced in association with mixed sneutrinos will be typically be accompanied by hard jets.  In this section, we consider the impact of mixed sneutrinos on jet-lepton invariant mass distributions, and find the following:
\begin{itemize}
\item{The decay chain $\tilde q \rightarrow (\chi^{\pm})   q\rightarrow (\tilde \nu_1 l) q$ leads to a prominant edge in this distribution, providing a potentially distinctive signature for mixed-sneutrino production. }
\item{A recently proposed mass-estimation method  \cite{Cheng:2007xv} can be used to probe the spectrum of the theory when applied to events involving this decay chain.  That method can thus be used to help distinguish mixed-sneutrino and MSSM scenarios, although in our implementation it does not reliably estimate the sneutrino mass.}
\end{itemize}

If  the sneutrino mixing angle $\theta$ is large enough, then chargino and neutralino decays directly
to $\snu_1$ can become important.  In fact, it is easy to imagine a situation in which 
$\chi_2^0 \rightarrow \snu_1 \nu$ and $\chi_1^\pm \rightarrow \snu_1 l$ both have nearly $100\%$
branching ratios.  To illustrate this with a concrete example, we take the high-scale parameters ${\tilde m}^2=(200$ GeV$)^2$, ${\tilde m}^2_{H_u,H,d}=0$ $M_{1/2}=300$ GeV, $\tan\beta =10$, $A_t=-500$ GeV, and
$A_b=A_\tau=0$.  We add to the resulting MSSM spectrum mixed sneutrinos with $\theta=0.2$ and $m_{\tilde \nu_1}=108$ GeV.   The superpartner spectrum for these parameters is given in table \ref{tab:spectrum2}.
\begin{table}[htdp]
{\begin{tabular}{||c||c||}
\hline
\hline
$m_{\tilde g}$ & 721 \\ 
\hline
$m_{{\tilde \chi}_2^\pm}$ & 536 \\
$m_{{\tilde \chi}_1^\pm}$ & 229 \\ 
\hline
$m_{{\tilde \chi}_4^0}$ & 536 \\
$m_{{\tilde \chi}_3^0}$ & 525 \\ 
$m_{{\tilde \chi}_2^0}$ & 229 \\
$m_{{\tilde \chi}_1^0}$ & 120 \\
\hline
\hline
\end{tabular} 
\hspace{0.3in}
\begin{tabular}{||c||c||}
\hline
\hline
$m_{{\tilde u}_L}$ & 684 \\ 
$m_{{\tilde u}_R}$ & 664 \\
\hline
$m_{{\tilde d}_L}$ & 688\\ 
$m_{{\tilde d}_R}$ & 663 \\
\hline
$m_{{\tilde t}_2}$ &682 \\ 
$m_{{\tilde t}_1}$ & 437 \\
\hline
$m_{{\tilde b}_2}$ &682 \\ 
$m_{{\tilde b}_1}$ & 663 \\
\hline
\hline
\end{tabular}
\hspace{0.3in}
\begin{tabular}{||c||c||}
\hline
\hline
$m_{{\tilde l}_L}$ & 281 \\ 
$m_{{\tilde l}_R}$ & 232 \\
\hline
$m_{{\tilde \tau}_2}$ & 291 \\ 
$m_{{\tilde \tau}_1}$ & 224 \\
\hline
$m_{\snu_2}$ &281 \\ 
$m_{\snu_1}$ & 108 \\
\hline
\hline
\end{tabular}
\hspace{0.3in}
\begin{tabular}{||c||c||}
\hline
\hline
$m_{H^\pm}$ & 561 \\ 
$m_{H}$ & 555 \\ 
$m_{A}$ & 555 \\ 
$m_{h}$ & 114 \\ 
\hline
\hline
\end{tabular}}
\caption{Superpartner and Higgs boson masses for the parameter point used to study the jet-lepton signature. All masses
are in GeV.}
\label{tab:spectrum2}
\end{table}%
From this table we see that the only two-body decays available to $\chi^\pm_1$ are to $\snu l$, $\neutone W$, and ${\tilde \tau}_1 \nu$.  The branching ratios for these final states are $95\%$, $4 \%$, and $<1\%$, respectively.
As before,
we make the simplifying assumption that all three generations of sterile sneutrinos have equal mixing angles with the active states.  If there is only one sterile sneutrino, which only mixes appreciably with $\snu_\tau$, then $\chi^\pm$ will
decay almost exclusively to $\snu_1 \tau$.  In this case detecting the chargino decays becomes more challenging.   

Although $\snu_1$ is lighter than $\neutone$ in the above spectrum, this ordering is not important for the signature
we are about to explore.  If we had taken $m_{\snu_1}=140$ GeV instead, we would still have $Br(\chi^\pm_1 \rightarrow \snu l)=93\%$.   So it is not essential for the following discussion that  $\snu_1$ is the LSP.  We should also note that, given
how dominant  $\chi^\pm_1 \rightarrow \snu l$ is for the parameters we've chosen, that decay mode still easily dominates for
more modest values of the mixing angle, $\theta \sim 0.1$.  So, signatures associated with $\chi^\pm_1 \rightarrow \snu l$ 
can exist for those smaller mixing angles as well.  

For the parameters chosen,  $\neuttwo$ also decays dominantly straight to $\snu_1$, with a branching ratio greater than $99\%$.   Because of the presence of $\snu_1$, both $\neutone$ {\em and} $\neuttwo$ appear as missing energy in the detector.   In particular, although there are leptons produced in chargino decays, there is no OSSF dilepton signal initiated
by $\neuttwo$ decays.

Chargino production via squark decay leads to the following sequence: $\tilde q \rightarrow (\chi^{\pm})   q\rightarrow (\tilde \nu_1 l) q$. One then expects to observe a kinematic endpoint in the jet-lepton invariant-mass distribution at
\begin{equation}
m_{ql}^{max} = m_{\tilde{q}} \sqrt{1 - (m_{\chi_1^\pm}/m_{\tilde{q}})^2} \sqrt{1-(m_{\snu_1}/m_{\chi_1^\pm})^2}.
\end{equation}
When we simulate events that include this sequence of decays using Pythia, the chargino is decayed isotropically in
its rest frame, and angular correlations between the lepton and jet are lost.  Taking $\chi_1^\pm$ to be pure charged wino,
which it nearly is for the chosen parameters, the squarks that can initiate this sequence are ${\tilde u}_L$, ${\tilde d}_L$, ${\tilde u}_L^*$, or ${\tilde d}_L^*$.   If these are produced with equal abundance, then the quark-lepton angular correlations  average
out to zero, even if we focus on a particular sign for the charge of the lepton.  In this case the fact that Pythia does not keep track of the angular correlations is not important, and the true $m_{ql}$ distribution looks the
same as the distribution for a 2-body/2-body sequence of decays with an intermediate scalar, shown as the dashed line of figure \ref{fig:22-23}. 

However, there is no reason to expect that ${\tilde u}_L$, ${\tilde d}_L$, ${\tilde u}_L^*$, and ${\tilde d}_L^*$ will be produced in 
equal abundance.  For example, for the sample point chosen above, the dominant SUSY production is squark+gluino.  The gluino
decays with roughly equal probabilities to all four of these possibilities, but the parton distribution functions dictate that the
squark produced is less likely to be  ${\tilde d}_L$ than ${\tilde u}_L$, and less likely still to be an anti-squark.   The sequence ${\tilde u}_L \rightarrow( \chi^{+})   d \rightarrow( \snu_1l^+) d$ gives a quark and a lepton with opposite helicities, while the sequence
 ${\tilde d}_L \rightarrow (\chi^{-} )  u \rightarrow( \snu_1^*l^-) u$ gives a quark and lepton with the same helicity.  So, the jet and lepton
 tend to be more back-to-back when produced by   ${\tilde d}_L$, and more in the same direction when produced by ${\tilde u}_L$.
 Given that ${\tilde u}_L$ is produced more abundantly than ${\tilde d}_L$, we should then expect that the combined $m_{ql}$ distribution
 will look somewhat softer than the dashed line of figure \ref{fig:22-23}, without the sharp edge.  On the other hand, if we focus on 
 jet -- $l^-$ invariant-masses, then we  should expect that  distribution to be even harder than for the case without angular correlations.
 This is because $l^-$ is produced in the sequences beginning with ${\tilde d}_L$ (which gives a quark and a lepton with the same helicity) or with ${\tilde u}_L^*$ (which gives an  antiquark and a lepton with the opposite helicity).  Assuming ${\tilde d}_L$ is produced more abundantly than ${\tilde u}_L^*$, the 
 jets and leptons will then tend to be more back-to-back on average.  
 
 Although these angular-correlation issues are important, we set them aside in what follows.
 The jet-lepton kinematic edge we identify below may be softened when both signs of lepton charge are allowed, but by requiring negatively charged leptons, an even harder distribution should result.

For the parameter point chosen above, we find a total SUSY production cross section of 20.4 pb.  We generate $\sim$ 160,000 events,
corresponding to $\sim$ 8 fb$^{-1}$, and keep events with the following characteristics:

\begin{itemize}

\item Exactly two jets with $p_T > 150$ GeV.

\item Exactly one isolated lepton with $p_T > 10$ GeV.

\item A transverse mass $m_T> 250$ GeV.

\item Missing transverse energy $E_T\!\!\!\!\!\! \slash \;\;> 250$ GeV.

\end{itemize}

\begin{figure}[h]
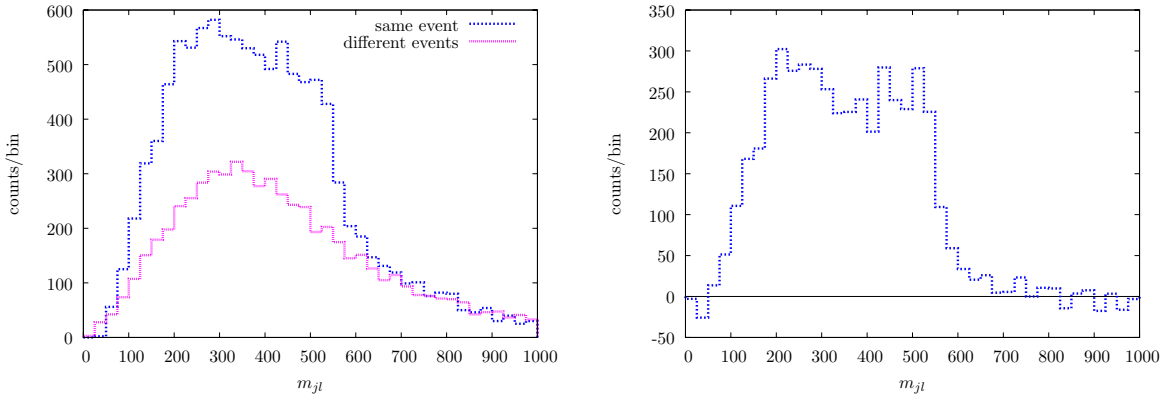
 
\centering
\scalebox{0.6}{\input{figjlcombined1120}}\quad  \scalebox{0.6}{\input{figjlsubtract1120}}
\caption{Left: invariant-mass distribution for  jets and leptons in the same event, and rescaled invariant-mass distribution for jets and leptons in different events.  Right: the subtracted distribution. }
\label{fig:jetlepton}
\end{figure}

We find that the number of $t{\overline t}$ events passing these cuts is  more than a factor of 20 smaller
than that from SUSY production. 
 Using Alpgen 2.12 \cite{Mangano:2002ea}, we estimate the $W$+jets background by obtaining $W$+2 jet events with a generator-level cut on the two jets, $p_T>100$ GeV.   We find that this source of background is even more suppressed than the $t{\overline t}$ background.

In figure \ref{fig:jetlepton} we show the jet-lepton invariant-mass distribution for the SUSY events, where for each event we include the two  invariant masses obtained by
pairing the isolated lepton with both of the hard jets.  A fairly steep drop-off in this distribution is seen near expected kinematic endpoint, which is at 568  GeV.  There is a large combinatorial
background associated with incorrect lepton-jet pairings, whose shape one  can attempt to guess  by looking at the invariant-mass distribution for  leptons-jet pairings from different events.  A similar method is used to study SUSY ditau signatures in \cite{CMS}, for example.  Whether the distribution for incorrect lepton-jet pairings from the same event and the distribution for lepton-jet pairings from different events are similar should obviously depend on the strength of the  correlations in the momenta of incorrectly paired jets and leptons from the same event.  Because the squarks will tend to be at least somewhat back-to-back,  one would expect correlations at some level.  
However, in the present example, the shape of the distribution for lepton-jet pairings from different events matches rather well with the shape of the same-event distribution
beyond the expected kinematic endpoint.  Rescaling the different-events distribution to match the same-event distribution at high invariant mass, and then subtracting that
rescaled distribution off, the drop-off near the kinematic endpoint becomes clearer, as shown in the second plot of figure \ref{fig:jetlepton} .  The bump in the distribution at relatively low invariant mass ($m_{jl} \lesssim 300$ GeV) is due to cascades involving $\tilde{t}_1$.

A jet-lepton invariant-mass distribution of the sort shown can also arise without mixing, from  production of ordinary sneutrinos.  Furthermore, a jet-lepton signal can be produced in other ways
in the MSSM.  A first example, relevant if the decay  $\tilde{\chi}_1^- \rightarrow \neutone W^-$ is dominant, is given by the sequence $\tilde{q} \rightarrow (\tilde{\chi}_1^-) q \rightarrow (\neutone W^-)q$, with $W^-$ decaying leptonically.   A second example is given by the decay
sequence $\tilde{q} \rightarrow (\tilde{\chi}_1^- )q \rightarrow ([{\tilde l}_L] {\overline \nu} )q \rightarrow ([\neutone l] {\overline \nu}) q $.  In both of these examples, the lepton and jet are accompanied by a neutrino, and so the kinematics are different than in case of sneutrino production.  One would expect the jet-lepton invariant-mass distributions to reflect these differences at some level.  In fact, for the second example one can  show that the distribution is softer than  for a 2-body/2-body sequence  with an intermediate scalar, regardless of the sign of the lepton considered.   The analysis for the first example is more complex, as there is no
intermediate scalar in the decay chain, but here there are other things to go on as well.
For example, one could look for events where the decay $\tilde{\chi}_1^\pm \rightarrow \neutone W^\pm$
occurs on both sides of the event, with one $W$ decaying leptonically and the other hadronically.
If a significant number of hadronic W's were reconstructed by looking for events of this type, the $\tilde{\chi}_1^\pm \rightarrow \neutone W^\pm$ interpretation would have to be favored over the  $\chi^{\pm}   \rightarrow \tilde \nu_1 l$ interpretation.   
Another potentially important difference is that one has flavor universality for the decays  $W^- \rightarrow l {\overline \nu}$, but
not necessarily for the decays $\chi^{-}   \rightarrow \tilde \nu_1^* l$.

Other observables may be useful for distinguishing production of mixed-sneutrinos 
and ordinary sneutrinos.  For example, for the parameter point considered above,
$\chi_2^0$ decays invisibly, and  the flavor-subtracted OSSF dilepton invariant-mass distribution 
has no particularly distinctive features.  On the MSSM side, for much of the parameter space for which  $\tilde{\chi}_1^-\rightarrow {\tilde \nu}^* l$  occurs,  $\tilde{\chi}_2^0\rightarrow {{\tilde l}_L}^* l$ also 
occurs, giving rise to an OSSF dilepton signature upon the subsequent decay $ {{\tilde l}_L}^* \rightarrow \neutone l^+$.  These decays tend to come along with each other because in the MSSM the masses of $\snu$ and ${\tilde l}_L$ are split only by electroweak symmetry breaking,
\begin{equation}
m^2_{{\tilde l}_L} - m^2_{\snu} \le m_W^2.
\end{equation}
Provided that $m_{\tilde{\chi}_1^\pm}$ is not much larger than $m_{\tilde{\chi}_2^0}$, the decay  $\tilde{\chi}_2^0\rightarrow {{\tilde l}_L}^* l$ thus tends to be  kinematically accessible when  $\tilde{\chi}_1^-\rightarrow {\tilde \nu}^* l$ is.  The essential point is that, because the mass of the mixed
sneutrino is not directly linked to the mass of charged slepton, it is easier than in the MSSM to have signals for sneutrino production in the absence of signals for charged slepton production.

One way to have $\tilde{\chi}_1^-\rightarrow {\tilde \nu}^* l$ without $\tilde{\chi}_2^0\rightarrow {{\tilde l}_L}^* l$ in the MSSM is to have a closely-spaced spectrum with $m_\snu < m_{\tilde{\chi}_1^\pm}$ and
$m_{{\tilde l}_L}> m_{\tilde{\chi}_2^0}$.   Mass measurements would clearly be helpful in distinguishing a mixed-sneutrino scenario scenario from this MSSM one.  For example, if it were established that the sneutrino-chargino mass splitting were quite large, it would disfavor the MSSM scenario just described.

\vskip 0.15in
\noindent {\it Mass Estimation}

Here we consider how the technique proposed by \cite{Cheng:2007xv} can be used to probe the mass spectrum.  The authors of that paper  consider a general situation in which 
 the sequence $Y \rightarrow l X$ and $X \rightarrow l' N$ occurs on both sides of an event.
The particle $N$  is invisible, so the final state topology involves four leptons and missing energy.  A typical
SUSY example of this situation has $Y = \neuttwo$, $X = {\tilde l}$, and $N=\neutone$.  
For the mixed-sneutrino scenario considered in this section, we have a large number of
events with $\tilde q \rightarrow \chi^\pm q \rightarrow  \tilde \nu l q$ on either side.  The event topology is thus quite  similar, but with two of the four leptons replaced with quarks. 

For a set of candidate values for the unknown masses $m_Y,m_X, m_N$, one can check whether the observed kinematics of a given event of this type are consistent with those values.   The procedure of \cite{Cheng:2007xv} is to fix two of the masses and  keep track of the number
of allowed events as the third mass is scanned.  A candidate value for the third mass is identified
by looking for a a dramatic feature in the in the resulting distribution, {\em e.g.} a sharp drop or peak in the number of allowed events. In our implementation we apply a smoothing procedure to the events distribution and identify the candidate mass as the point where the second derivative of the distribution is minimized.

These steps are then iterated -- the third mass is fixed at its new candidate value while the
first mass is scanned, and so on.  It was found in \cite{Cheng:2007xv} that this procedure does not converge, and it was suggested that  the actual masses can be estimated as the ones that give a global peak in the number of consistent events as the iterations are performed.  In our implementation we find that for some of the events samples the candidate masses quickly settle near a final value, while for others the candidate masses continue to jump around indefinitely.  Even in the case where the candidate masses continue to jump around,  they at least wind up in stable ``orbits'' after a sufficient number of iterations.  We take the average values of these orbits as the mass estimates for a given event sample.  

Although we do not find that our implementation of this procedure leads to a reliable estimate of the
sneutrino mass, it does give a reliable estimate of the chargino-sneutrino mass splitting.  
We first apply the technique to the same parameter point described above,  taking the most  optimistic case where the chargino decays produce electrons and muons and not taus.  We select events with exactly two jets with $p_T > 150$ GeV,
and  exactly two leptons with $p_T > 10$ GeV.  The leptons are required to have the same sign in order to suppress standard model backgrounds.
We find that the efficiencies for passing these cuts are  1.4\% for the SUSY sample 
and  $1.3 \times 10^{-5}$ for the $t{\overline t}$ sample, giving a ratio of SUSY events to $t{\overline t}$ events of over 20. 
 More problematic than the $t{\overline t}$ background is the background from SUSY events
 in which the selected jets and leptons do not come from the desired decay chains.  
Using 12 sets of 1,000 events, we obtain 12 estimates for  $(m_{\tilde q},m_{\tilde{\chi}_1^\pm} ,m_{\snu_1} )$.  Combined,
these event samples correspond to an integrated luminosity of roughly 42 fb$^{-1}$.  For each event sample, we take the average values of the candidate masses at large iteration number as the mass estimates for that sample.  Averaging {\em these} estimates gives
\begin{eqnarray*}
(m_{\tilde q},m_{\tilde{\chi}_1^\pm} ,m_{\snu_1} ) &  =  & (688 \pm 33\;{\rm GeV},239\pm 27\;{\rm GeV},110 \pm 30\;{\rm GeV})\\
m_{\tilde{\chi}_1^\pm} - m_{\snu_1}  & = & 129 \pm 7\;{\rm GeV},
\end{eqnarray*}
compared with the actual values, $(m_{\tilde q},m_{\tilde{\chi}_1^\pm} ,m_{\snu_1} ) =(684-688 \;{\rm GeV},229 \;{\rm GeV},108 \;{\rm GeV})$
and $m_{\tilde{\chi}_1^\pm} - m_{\snu_1}  = 121 \;{\rm GeV}$. 
If the chargino decays produce $e$, $\mu$, and $\tau$ with the same probability, the
resolution worsens somewhat. 
Performing the same procedure on 10 sets of 1,000 events, we obtain 
\begin{eqnarray*}
(m_{\tilde q},m_{\tilde{\chi}_1^\pm} ,m_{\snu_1} ) &  =  & (720 \pm 40\;{\rm GeV}, 263\pm 35\;{\rm GeV},140 \pm 42\;{\rm GeV})\\
m_{\tilde{\chi}_1^\pm} - m_{\snu_1}  & = & 123 \pm 10\;{\rm GeV}.
\end{eqnarray*}
If we now redo the same analysis with the same parameters except with  the sneutrino mass increased to 142 GeV,
we obtain $m_{\snu_1} =116 \pm 22\;{\rm GeV}$ and $m_{\tilde{\chi}_1^\pm} - m_{\snu_1}   = 92 \pm 7\;{\rm GeV}$ (versus an actual splitting of 87 GeV). Increasing the sneutrino mass further to 185 GeV gives $m_{\snu_1} =124 \pm 23\;{\rm GeV}$ and $m_{\tilde{\chi}_1^\pm} - m_{\snu_1}   = 48 \pm 2\;{\rm GeV}$ (versus an actual splitting of 44 GeV). 

These results show that although the sneutrino mass estimates do not follow the actual values closely, the estimates of the chargino-sneutrino mass splitting do. So, this analysis  can be used to find  evidence against  a closely-spaced spectrum of the type described above.  A more sophisticated implementation of this mass-estimation method, or different methods such as those proposed in \cite{Cho:2007qv,Lester:2007fq} may do a better job at estimating the
sneutrino mass itself.

\subsection{Associated production of Z/h with a lepton}
\label{subsection:zh}

For some regions of parameter space, the heavier sneutrino would be produced at the LHC.  In particular, if a chargino with significant charged-wino component is heavier than the heavier sneutrino, it can decay to  that  sneutrino and a charged lepton.  The question is then how the
 heavier sneutrino decays. One possibility is  $\snu_2 \rightarrow \neutone \nu$, which is just what one  might expect in the MSSM.  Here we consider a more distinctive scenario, in which the splitting with the lighter sneutrino and the sneutrino
 mixing angle are both large enough that the branching ratios for $\snu_2 \rightarrow \snu_1 Z$ and $\snu_2 \rightarrow \snu_1 h$ are both significant.  In this case we  find
 \begin{itemize}
 \item{Cascade production of $Z$ bosons leads to a clean signature if the sneutrino contains  $e$ or $\mu$ flavor.  The shape of the trilepton invariant mass distribution can in principle distinguish this signature from MSSM signatures involving $Z$ bosons.}
  \item{If the sneutrino is dominantly $\tau$ flavored, a much larger integrated luminosity of the LHC is necessary to see cascade $Z$ production}
 \item{Cascade production of Higgs bosons can also lead to distinctive signatures, both through the $h\rightarrow b \bar b$ and $h \rightarrow \gamma \gamma$ decay channels.}
 \end{itemize}

As an example of a point in parameter space with these interesting heavy-sneutrino decays, consider the weak-scale values shown in table \ref{tab:h/z_params}, and the resulting mass spectrum of
table \ref{tab:h/z_masses}. 
\begin{table}[htdp]
{\begin{tabular}{||c|c|c||c|c|c||c|c||c|c|c||c|c||}
\hline
\hline
$\tan\beta$  & $\mu$  & $m_A$ & $M_1$ & $M_2$ & $M_3$ & $A_t$& $A_{b,\tau}$ &  ${\tilde m}_{Q,u,d}^2$ & ${\tilde m}_L^2$ & ${\tilde m}_e^2$ & $m_{\snu_1}$& $\theta$   \\ 
\hline
\hline
$10$  & $600$  & $350$ & $200$ & $500$ & $700$ & $-800$& $0$ &  $(600)^2$ & $(300)^2$ & $(250)^2$ & $82$& $0.2$   \\ 
\hline
\hline
\end{tabular}} 
\caption{Parameters chosen to study $\snu_2 \rightarrow \snu_1 Z$ and $\snu_2 \rightarrow \snu_1 h$ signatures.  All masses are in GeV.}
\label{tab:h/z_params}
\end{table}
\begin{table}[htdp]
{\begin{tabular}{||c||c||}
\hline
\hline
$m_{\tilde g}$ & 721 \\ 
\hline
$m_{{\tilde \chi}_2^\pm}$ & 629 \\
$m_{{\tilde \chi}_1^\pm}$ & 474 \\ 
\hline
$m_{{\tilde \chi}_4^0}$ & 630\\
$m_{{\tilde \chi}_3^0}$ & 601 \\ 
$m_{{\tilde \chi}_2^0}$ & 474 \\
$m_{{\tilde \chi}_1^0}$ & 196 \\
\hline
\hline
\end{tabular}
\hspace{0.3in}
\begin{tabular}{||c||c||}
\hline
\hline
$m_{{\tilde u}_L}$ & 623 \\ 
$m_{{\tilde u}_R}$ & 624 \\
\hline
$m_{{\tilde d}_L}$ & 628\\ 
$m_{{\tilde d}_R}$ & 626 \\
\hline
$m_{{\tilde t}_2}$ &734 \\ 
$m_{{\tilde t}_1}$ & 524 \\
\hline
$m_{{\tilde b}_2}$ &639 \\ 
$m_{{\tilde b}_1}$ & 615 \\
\hline
\hline
\end{tabular}
\hspace{0.3in}
\begin{tabular}{||c||c||}
\hline
\hline
$m_{{\tilde l}_L}$ & 303 \\ 
$m_{{\tilde l}_R}$ & 254 \\
\hline
$m_{{\tilde \tau}_2}$ & 309 \\ 
$m_{{\tilde \tau}_1}$ & 249 \\
\hline
$m_{\snu_2}$ &299 \\ 
$m_{\snu_1}$ & 82 \\
\hline
\hline
\end{tabular}
\hspace{0.3in}
\begin{tabular}{||c||c||}
\hline
\hline
$m_{H^\pm}$ & 359 \\ 
$m_{H}$ & 351 \\ 
$m_{A}$ & 351 \\ 
$m_{h}$ & 114 \\ 
\hline
\hline
\end{tabular}}
\caption{Superpartner and Higgs boson masses for the parameter point of table \ref{tab:h/z_params}}
\label{tab:h/z_masses}
\end{table}%
We find that the SUSY production cross-section for this spectrum is 24 pb, and
the relevant branching ratios for the signals that interest us here are 
$Br(\chi^+_1 \rightarrow \snu_2 l)=32\%$ ($l=e,\mu$), $Br(\snu_2 \rightarrow \snu_1 Z)=37\%$, and $Br(\snu_2 \rightarrow \snu_1 h)=37\%$.  These decays lead to a trilepton signature for
the case of $Z$ production, and $l b\bar b$ and $l \gamma \gamma$ signatures for the case of Higgs production.  

\begin{figure}[h]
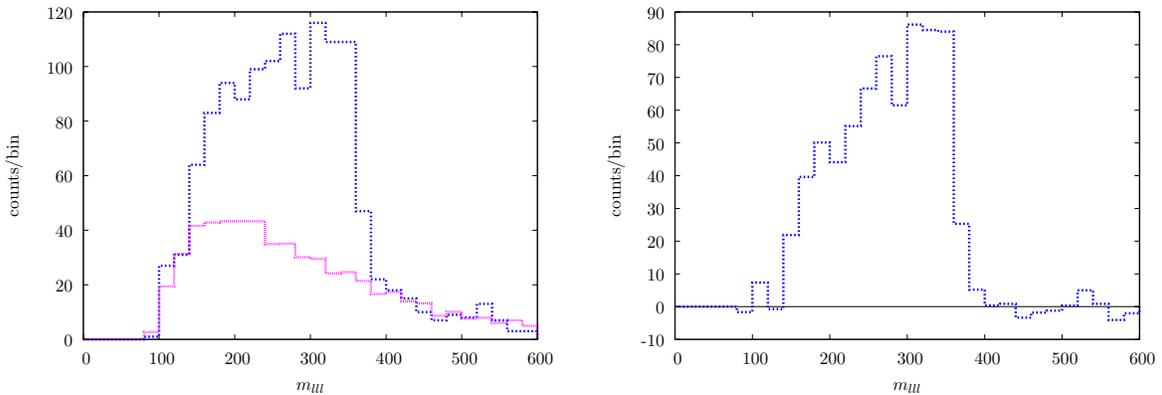
 
\centering
\scalebox{0.6}{\input{figlzcombined}}\quad  \scalebox{0.6}{\input{figlzsubtract}}
\caption{Left: invariant-mass distribution for  trileptons in the same event, and rescaled invariant-mass distribution for Z candidates and leptons in different events.  Right: the subtracted distribution. }
\label{fig:lz}
\end{figure}

To explore these signatures, we generate $~640$k SUSY events, corresponding to about
27 fb$^{-1}$ of integrated luminosity.   We select events with the following properties to 
study the lepton-$Z$ signature:
\begin{itemize}

\item Three isolated leptons with $p_T > 10$ GeV.  Two of these must be of opposite sign and
same flavor, with $|m_{l^+l^-}-m_Z|<10$ GeV.

\item  $\sum p_T>800$ GeV, where  the sum is over jets with $p_T>20$ GeV,
leptons and photons with $p_T>10$ GeV, and missing $p_T$.

\end{itemize}
These cuts  select 1,323  SUSY events.  We find the background from 
$t {\overline t}$ and $WZ$ production (also generated using Pythia) are negligible by comparison.  The trilepton invariant mass distribution for events passing the cuts is shown
in the first plot of figure \ref{fig:lz}
The distribution shows a steep drop-off near the expected endpoint at 364 GeV.  Along the lines of what was done for the jet-lepton mass distribution, one
can attempt to subtract off the SUSY background by guessing that the shape of the SUSY background contribution to the distribution should be similar to to the distribution obtained by matching the  $Z$ candidates from one event with leptons from different events.  Subtracting off
this distribution gives the distribution shown in the second plot of figure \ref{fig:lz}.  
A lower kinematic endpoint should in principle be observed at around 148 GeV.  With
greater statistics it is possible that a second edge near this value would become clear in
the subtracted distribution.  

$Z$ bosons can also be produced in association with leptons in the MSSM,  in processes such as
$\tilde{\chi}^-_2 \rightarrow (\tilde{\chi}^-_1 )Z \rightarrow( \snu^* l) Z$.  In principle, this MSSM process
is distinguishable from the mixed-sneutrino signature considered above by examining the $Z$-$l$
invariant-mass distribution.  In the mixed-sneutrino case, the intermediate $\snu_2$ decays isotropically in its rest frame.   The amplitude-squared for the chargino decay into $\snu_1lZ$ is thus constant,
and the $m_{Zl}$ distribution is trapezoidal, rising linearly between  lower and upper kinematic edges.
In the MSSM sequence $\tilde{\chi}^-_2 \rightarrow (\tilde{\chi}^-_1) Z \rightarrow (\snu^* l) Z$, there is
no intermediate scalar, and there will generically  be some dependence of the amplitude-squared on $m_{Zl}$.  The chargino-chargino-$Z$  coupling can be written as
\begin{equation}
{\overline \chi_2^-} \gamma ^\mu (c_L P_L + c_R P_R) \chi_1^-,
\end{equation}
where the couplings $c_L$  and $c_R$  are determined by the mixing in the chargino sector.  
These couplings are typically not equal.  In the narrow-width approximation, the amplitude-squared for the decay 
$\tilde{\chi}^-_2 \rightarrow \snu^* l Z$ takes the form $\alpha+\beta (|c_L|^2 -|c_R|^2) m_{lZ}^2$, where $\alpha$ and $\beta$ are constants.  Provided the chargino-$Z$ coupling is indeed chiral, the invariant mass distributions thus differ in the MSSM and mixed-sneutrino cases.  A similar issue was raised in considering the possible production of $Z$ bosons in top-squark decays in \cite{Perelstein:2007nx}.

If only the $\snu_\tau$ mixes with a sterile neutrino, the signature becomes much less clean than what we have considered.  In this
case $\snu_{e,\mu}$ decay straight to $\neutone \nu$, and so one is forced to look for signals from 
associated $\tau-Z$ production.  To explore these we use a much larger sample of SUSY events
corresponding to roughly 170 fb$^{-1}$, and impose the same cuts as before, except that his time
we require a reconstructed $\tau$ and two opposite sign, same flavor leptons that reconstruct a $Z$.
\begin{figure}[h]
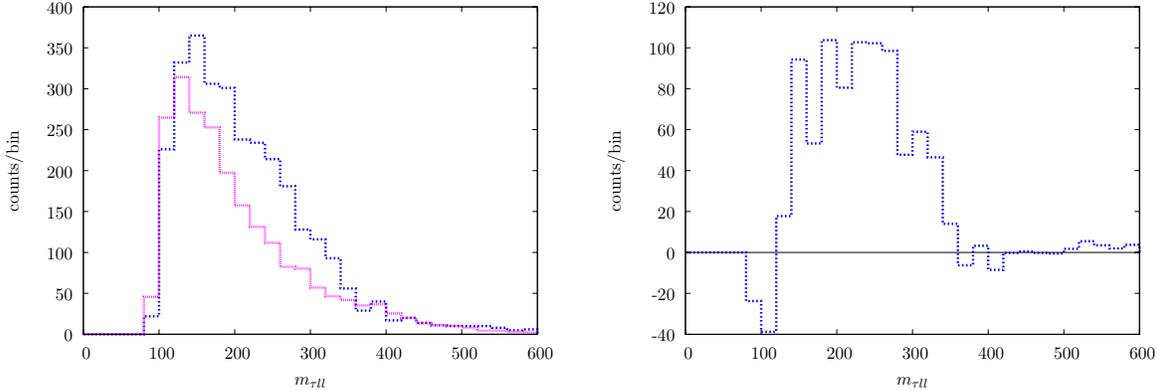
 
\centering
\scalebox{0.6}{\input{figtauZ}}\quad  \scalebox{0.6}{\input{figtauZsubtract}}
\caption{Left: $\tau ll$ invariant-mass distribution for  $\tau$'s and $Z$ candidates in the same event, and rescaled invariant-mass distribution for $\tau$'s and $Z$ candidates in different events.  Right: the subtracted distribution. }
\label{fig:tauZ}
\end{figure}
After these cuts, we are left with 3,020 SUSY events.  We estimate that the $t {\overline t}$ background
gives  fewer events by  a factor of 4.8.  In our plots we will not include the $t {\overline t}$ contribution, as we have not generated a large enough $t {\overline t}$ sample; it is likely that more carefully chosen cuts
can improve the quoted signal to background ratio. 

In the first plot of figure \ref{fig:tauZ} we show the $\tau ll$ invariant-mass distribution for events passing the cuts, along with the (rescaled) distribution obtained by matching $Z$ candidates from one event with $\tau$'s from a different event.  The second plot of the
same figure shows the subtracted distribution, which does have something that looks like an endpoint.  
The excess of events in the different-event distribution at low invariant mass is presumably due to
the fact that the isolation criteria for $\tau$'s and leptons in the same event for that are not enforced for 
$\tau$'s and leptons in different events.  

\begin{figure}[h]
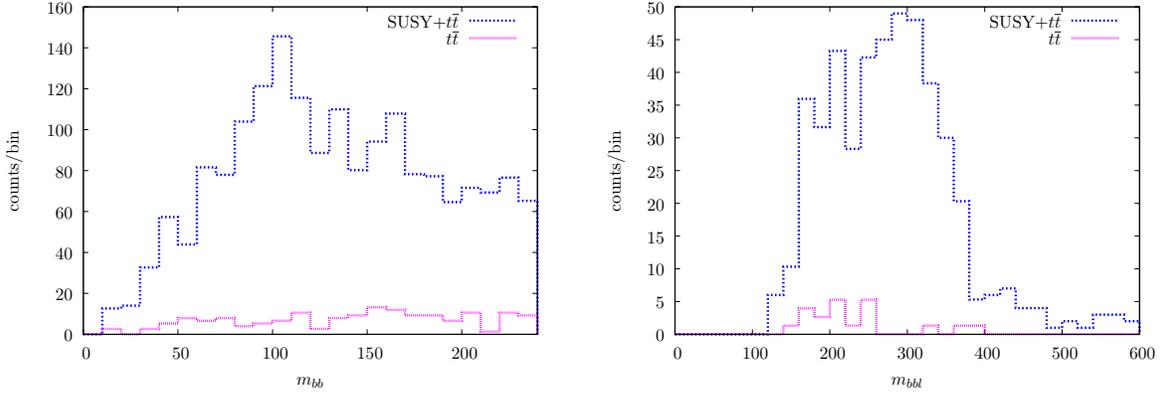
 
\centering
\scalebox{0.6}{\input{figbb}}\quad  \scalebox{0.6}{\input{figbbl}}
\caption{Left: invariant-mass distribution for pairs of $b$-tagged jets.  Right: for events in the peak of the $m_{bb}$ distribution,  invariant-mass distribution obtained by pairing Higgs candidates and leptons. }
\label{fig:lh}
\end{figure}

To study the lepton-Higgs signature, we first select events with the following
characteristics:
\begin{itemize}

\item Two $b$-tagged jets with $p_T > 20$ GeV.

\item Exactly one isolated lepton with $p_T > 10$ GeV.

\item A transverse mass $m_T> 200$ GeV.

\item Missing transverse energy $E_T\!\!\!\!\!\! \slash \;\;> 200$ GeV.

\item  $\sum p_T>800$ GeV, where  the sum is over jets with $p_T>20$ GeV,
leptons and photons with $p_T>10$ GeV, and missing $p_T$.

\end{itemize}
Working with the same 27 fb$^{-1}$ sample used to analyze the lepton-$Z$ signature, a total of 2,432 SUSY events survive these cuts, a factor of eight larger than
the number of $t {\overline t}$ events that survive.  The invariant mass distribution for
events passing these cuts is shown in the first plot of  figure \ref{fig:lh}.  
A peak, although
not a clean one, is evident at around 100 GeV, somewhat below the actual
Higgs mass of 114 GeV.  We keep events events with $|m_{bb}-100\;{\rm GeV}|< 20$ GeV,
and the $m_{bbl}$ invariant-mass distribution for those events is shown in the second
plot of figure \ref{fig:lh}.  The distribution shows a significant drop-off near the expected
endpoint, at 369 GeV.  

For the parameter point chosen, the rate of Higgs production is large enough that
it the Higgs can also be seen through its decays to photons.  For this analysis, we impose
the same cuts as for the $bbl$ analysis, except that we require two photons with $p_T > 10$ GeV
instead of two $b$-tagged jets.  
\begin{figure}[h]
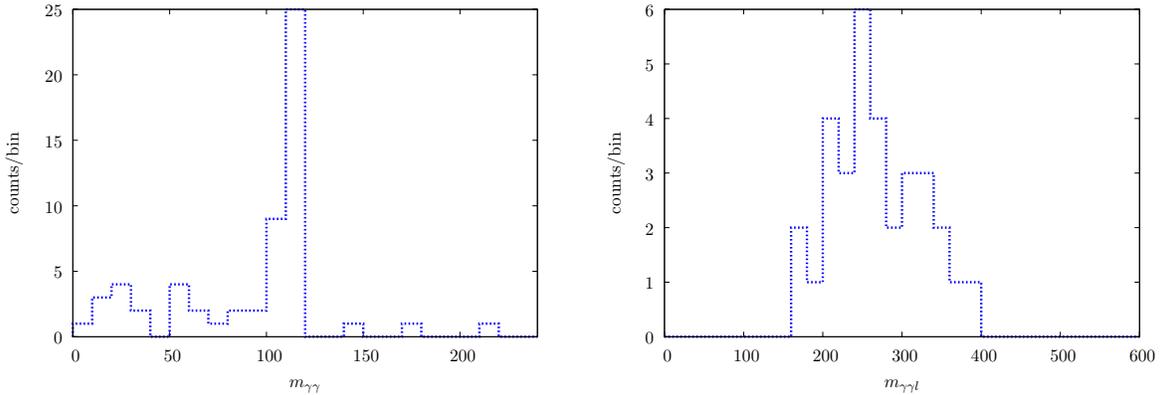
 
\centering
\scalebox{0.6}{\input{figgamgam}}\quad  \scalebox{0.6}{\input{figgamgaml}}
\caption{Left: invariant-mass distribution for pairs of photons.  Right: for events in the peak of the $m_{\gamma \gamma}$ distribution,  invariant-mass distribution obtained by pairing Higgs candidates and leptons. }
\label{fig:gamgam}
\end{figure}
A total of 60 SUSY events pass these cuts.  We have checked that the backgrounds  from $t \overline{t}$ production and $W \gamma \gamma+$jets  (estimated using Alpgen) are negligible. (In fact, it is possible that it would be advantageous to relax the kinematic cuts.)
 The $\gamma \gamma$ invariant mass distribution for
events passing these cuts is shown in the first plot of  figure \ref{fig:gamgam}, and has an extremely clear peak.
The $m_{\gamma \gamma l}$  distribution for  events  in this peak is shown in the second
plot of figure \ref{fig:gamgam}; it  falls below $\sim$ 400 GeV as one would expect, but  more statistics would be required to learn much from this distribution.

\section{Conclusions}
In this paper we have explored the cosmology and collider phenomenology of mixed sneutrinos.
Recent progress made by direct-detection experiments makes mixed-sneutrino dark matter quite constrained, but still viable.
In the absence of lepton-number violation,  parameter regions that give an appropriate relic abundance while evading direct-detection constraints include (1) light sneutrinos, $m_{\snu_1} < 10  \;{\rm GeV}$, with relatively light gauginos and a  rather large mixing angle, $\theta \sim 0.3$,  (2) small mixing angles, $\theta \lesssim 0.07$, with $m_{\snu_1}$ near the Higgs funnel, or (3) small mixing angles with $m_{\snu_1}$ above threshold for annihilation to W pairs, for large values of heavier sneutrino masses, $m_{\snu_2} \gtrsim 500 \;{\rm GeV}$.

Lepton-number violation in the sneutrino mass-sqaured matrix can suppress scattering of $\snu_1$ off of nuclei via $Z$ exchange, making somewhat
larger mixing angles viable.  However, this lepton-number violation  produces radiative contributions to neutrino mass.  If the lepton-number violation is large enough to dramatically suppress the elastic scattering via  $Z$ exchange, the radiatively generated mass tends to approach or exceed the upper bound  from cosmology.  In principle one can suppress this radiatively generated neutrino mass by
making the gauginos heavy, while still achieving a realistic relic abundance through annihilations mediated by Higgs and $Z$ exchange, or by making the gauginos Dirac.

We have studied LHC signatures of mixed sneutrinos in general, without requiring the mixed sneutrino responsible for the signal to be the dark matter.   If the mixed-sneutrino  is the LSP, with a very small mixing angle,
then the NLSP will will typically be the only particle that decays to it with a large branching ratio.  If this particle is the lightest neutralino, then the only effect of the mixed sneutrino is to alter the connection between collider physics and cosmology.
If the NLSP is instead a right-handed slepton one expects an unusually large lepton multiplicity in the SUSY signal.    Moreover,  decays of the lightest neutralino can lead to an interesting opposite-sign, same-flavor (OSSF) dilepton signature.
Since the dilepton signature arises  from a two-body decay followed by a three-body decay, the shape of the dilepton invariant-mass distribution is significantly different from the sequence of two-body decays
$\chi_2^0 \rightarrow (\tilde l)  l^+ \rightarrow (\chi_1^0  l^-) l^+ $.  It can also be distinguished from that
arising from a three-body decay $\chi_2^0 \rightarrow \chi_1^0  l^+ l^-$, depending on factors such as the observed kinematic endpoint of the distribution.  If the lightest neutralino decays to  $\tilde{\tau_1}$ rather than  ${\tilde e}_R$ or ${\tilde \mu}_R$, the experimental signatures become more difficult
to extract, but there is still the possibility of observing a large excess of events with taus produced in association with opposite-sign leptons, where the leptons come from $W$ bosons produced in the
decay ${\tilde \tau}_1 \rightarrow \snu_1 W$.

For larger mixing angles, it is possible that the mixed sneutrinos will be produced copiously at the LHC through chargino decays, or even from the decays of heavier sneutrinos.  In the first case, the sequence
$\tilde q \rightarrow( \chi^{\pm} )  q\rightarrow( \tilde \nu_1 l) q$ gives rise to a kinematic edge in the jet-lepton invariant-mass distribution.   For broad regions of parameter space, not only $ \rightarrow \chi^{\pm} $ but also $\neuttwo$ decays dominantly to $\snu_1$, and in these regions the jet-lepton signature is present in the absence of an OSSF dilepton signature.  This situation can also arise due to ordinary sneutrino production in the MSSM, but we have shown that mass-estimation methods may be helpful for distinguishing the mixed-sneutrino and MSSM scenarios, due to the fact that the sneutrino - charged slepton mass splitting is  an electroweak symmetry breaking effect in the MSSM, but not in the mixed-sneutrino case.

Finally, if the predominantly left-handed sneutrinos $\snu_2$ are produced at the LHC, the decays
 $\snu_2 \rightarrow \snu_1 Z$ and  $\snu_2 \rightarrow \snu_1 h$ may be important if  kinematically
accessible.  Events with $\snu_2$ is produced from chargino decay may then have Higgs or $Z$ bosons produced in association with leptons.  In this case, a distinctive $b {\overline b} l$ invariant mass distribution  can arise from Higgs production, and a distinctive trilepton invariant-mass distribution can arise from $Z$ production.  Because these decay chains feature an intermediate scalar, $\snu_2$,
they can in principle be distinguished from MSSM decay chains such as $\tilde{\chi}^-_2 \rightarrow( \tilde{\chi}^-_1) Z \rightarrow (\snu^* l) Z$, where the chargino-chargino-$Z$ coupling is in general chiral.
We leave a detailed analysis of this issue for future work.

\section{Acknowledgments}
The work of ZT and DTS was supported by NSF grant 0555421 and a Research Corporation Cottrell College Science Award.  NW was supported by NSF CAREER grant PHY-0449818 and DOE grant number DE-FG02-06ER41417.

\bibliography{tuckersnu}
\bibliographystyle{apsrev}
\end{document}